\documentclass[traditabstract]{aa}
\usepackage{amssymb,amsmath}
\usepackage{graphicx}
\usepackage[varg]{txfonts}
\usepackage{color}
\usepackage{natbib}

\begin{document}

\title{Phase-space structure analysis of self-gravitating collisionless spherical systems}

\author{A. Halle \inst{1,2}\thanks{E-mail:halle@iap.fr} \and S. Colombi\inst{1} \and S. Peirani \inst{1,3}}

\institute{UPMC-CNRS, UMR7095, Institut d'Astrophysique de Paris, 98 bis boulevard Arago, 75014 Paris, France \and Max Planck Institut f\"ur Astrophysik, Karl-Schwarzschild-Strasse 1,  D-85741 Garching bei M\"unchen, Germany \and Universit\'e C\^ote d’Azur, Observatoire de la C\^ote d’Azur, CNRS, Laboratoire Lagrange, Bd de l'Observatoire, CS 34229, 06304 Nice Cedex 4, France}

\date{Received / Accepted}

\abstract{In the mean field limit, isolated gravitational systems often evolve towards a steady state through a violent relaxation phase. One question is to understand the nature of this relaxation phase, in particular the role of radial instabilities in the establishment/destruction of the steady profile.  Here, through a detailed phase-space analysis based both on a spherical Vlasov solver, a shell code and a $N$-body code, we revisit the evolution of collisionless self-gravitating spherical systems with initial power-law density profiles $\rho(r) \propto r^n$,  $0 \leq n \leq-1.5$, and Gaussian velocity dispersion. Two sub-classes of models are considered, with initial virial ratios $\eta=0.5$ (``warm'') and $\eta=0.1$ (``cool''). Thanks to the numerical techniques used and the high resolution of the simulations, our numerical analyses are able, for the first time, to show the clear separation between two or three well known dynamical phases: (i) the establishment of a spherical quasi-steady state through a violent relaxation phase during which the phase-space density displays a smooth spiral structure presenting a morphology consistent with predictions from self-similar dynamics, (ii) a quasi-steady state phase during which radial instabilities can take place at small scales and destroy the spiral structure but do not change quantitatively the properties of the phase-space distribution at the coarse grained level and (iii) relaxation to non spherical state due to radial orbit instabilities for $n \leq -1$ in the cool case.}


\maketitle

\section{Introduction}
Dark matter in the Universe and stars in galaxies behave like a self-gravitating collisionless fluid of which the dynamics can be described by the Vlasov-Poisson system : 
\begin{eqnarray}
\dfrac{\partial f}{\partial t} + \mathbf{v} \cdot  \dfrac{\partial f}{\partial \mathbf{x}} -\dfrac{\partial \phi}{\partial \mathbf{x}} \cdot \dfrac{\partial f}{\partial \mathbf{v}} = 0, \label{eq:vlagen1} \\
\vartriangle \phi = 4 \pi G \rho = 4 \pi G \int f(\mathbf{x}, \mathbf{v}, t) \mathrm{d} \mathbf{v}, \label{eq:vlagen2}
\end{eqnarray}
where $f(\mathbf{x}, \mathbf{v}, t)$ is the phase-space density of the fluid at position $\mathbf{x}$, velocity $\mathbf{v}$ and time $t$, $\rho$ is the mass density and $\phi$ is the gravitational potential.

A major issue when considering the dynamics of gravitational systems such as dark matter halos, elliptical galaxies or star clusters in the non collisional regime is to understand the main processes underlying the creation of the quasi-stationary states that build up after a number of dynamical times, for instance the universal profiles of dark matter halos \citep{NFW1,NFW2}. 

One way to relate initial to quasi-equilibrium state is to assume that the system reaches some maximum entropy state after a {\em violent relaxation} phase with strong mixing \citep{LyndenBell}. However, the maximum entropy approach is at best partly successful \citep[see, e.g.][and references therein]{Yamashiro1992,Arad2005, Arad2005b,Yamaguchi2008,Joyce2011} and the only way to improve the results is to introduce additional constraints and ad hoc ingredients \citep[see, e.g.][]{Hjorth2010,Pontzen2013,Carron2013}. Indeed, relaxation might be incomplete, or maximum entropy state might not even exist, although one can generalize the concept of entropy by introducing the more general concept of H-functions \citep{Tremaine1986}, for which there exist stationary points related to actual stable equilibria even if entropy maximum does not exist. 

Another popular alternative to try understanding the establishment of quasi-stationary profiles consists in investigating the subspace of self-similar solutions \citep[see, e.g.,][but this list is far from exhaustive]{fillmore84,bertschinger85,Henriksen1995,sikivie97,mohayaee06,alard13}. While it is difficult to actually demonstrate the onset of self-similarity, it seems to be a natural outcome of gravitational dynamics, at least in the absence of any characteristic scale. 
Although they exist as well in the warm case \citep[see, e.g.][]{Henriksen1995}, self-similar solutions have been mostly studied in the cold case, for which the phase-space distribution function is of zero initial velocity dispersion. In this configuration, a $D$-dimensional phase-space sheet evolves in $2D$ phase-space and builds up a spiral pattern.

The above approaches, along with perturbation theory in cosmological systems \citep[see, e.g.][]{ptreview}, provide some partial analytical framework to study the Vlasov-Poisson system. However, in general, these equations usually require a numerical approach, which consists in decomposing the phase-space distribution function on an ensemble of macro-particles interacting with one another with a softened gravitational force \citep[see, e.g.][for reviews on the subject]{1988csup.book.....H,Bertschinger1998,Colombi2001,Dolag2008,Dehnen2011}. An alternative way, easily tractable in a small number of dimensions or for systems with a high level of symmetry, consists in using direct Vlasov solvers where the phase-space distribution function is generally sampled on a Eulerian mesh. Most of the direct Vlasov solvers have been developed in plasma physics and are of {\em semi-Lagrangian} nature. They exploit directly Liouville theorem, namely that the phase-space density is constant along characteristics. In the standard semi-Lagrangian scheme, a test particle is associated to each grid site where $f$ has to be calculated. This particle is followed back in time during a time-step and the value of $f$ is given by the interpolation of the phase-space density at previous time-step at the root of the trajectory. In the seminal implementation by \citet{cheng76}, this is performed in a split fashion between velocities and positions. Many improvements and modifications have been added over time to the splitting algorithm of Cheng \& Knorr \citep[see, e.g., the extensive review in the introduction of ][]{Sousbie2016}, mainly by plasma physicists. The splitting scheme was first applied to astrophysical systems by \citet{Fujiwara1981}, \citet{Nishida1981} and \citet{Watanabe1981}. In the classical implementation that we shall use below, re-interpolation of the phase-space distribution function is performed at each time-step using a third order spline.  

In all the cases, validating the results obtained from numerical resolution of Vlasov-Poisson equations remains difficult, particularly if one aims to remain in the mean field limit. In particular, $N$-body results are often debated. For instance, the close $N$-body encounters and collective effects due to particle shot noise can have some dramatic, possibly cumulative effects \citep[see, e.g.][]{Aarseth1988,Kandrup1991,Boily2002,binney04,Joyce2009,colombi15,Beraldo2017,Romero2018}, particularly when initial conditions are cold or close to cold \citep[see, e.g.,][]{Melott1997,Melott2007}.  While Vlasov codes do not use particles, they are still subject to non trivial numerical effects, because a phase-space grid still remains a discrete representation of the system \citep[see, e.g.][hereafter C15]{colombi15}. But since the numerical implementation is still different from $N$-body codes, a comparison between Vlasov and $N$-body codes seems appropriate and timely, especially when trying to analyse in detail the quasi-stationary state reached in the fluid limit by gravitational systems. 

One question indeed remains open. What are the main processes involved in the violent relaxation phase leading to a quasi-stationary state? We propose here to approach this question by studying the evolution of a number of initially spherical systems with various initial density profiles and velocity dispersions, focusing on the phase-space structure. Spherical symmetry will allow us to compare high resolution Vlasov simulations to  $N$-body simulations. Our analyses will focus on the detailed structure of the phase-space distribution function and comparisons with predictions from self-similarity.
 
The advantage of systems with initial spherical symmetry is that they have been studied in great detail in the literature both from the theoretical and the numerical points of view. One major question for instance concerns the role of radial instabilities and radial orbit instability in the establishment of the quasi-steady state observed after violent relaxation. Some spherical equilibria or quasi-steady states are known to be unstable to radial perturbations \citep[see, e.g.][]{Henon73, henriksen97} as well as angular perturbations that translate into radial orbit instability as well studied in the literature \citep[see, e.g.][]{Polyachenko1981,Merritt1985,Barnes1986,Cannizzo92,Barnes2009, Marechal2011, Vogelsberger2011,Polyachenko2015}. These perturbations are usually induced by shot noise due to the discrete nature of the distribution of stars or particles in the system, which directly relates to the discussion above about the validity of numerical simulations. Here, it is interesting to see what happens in the mean field limit, or at least in a regime that tries to approach it by using a direct Vlasov code and $N$-body simulations with a very large number of particles. 

More specifically, assuming $G=1$ and following the footsteps of \cite{Burkert1990}, \cite{Hozumi96} and \cite{hozumi00}, we perform a number of controlled numerical experiments of unity total mass systems, initially spherical with a power-law density profile and a Gaussian isotropic velocity dispersion: 
\begin{eqnarray}
f(\mathbf{r},\mathbf{v}) &=& \frac{\rho_0 (r)}{(2\pi \sigma_r^2)^{3/2}} \exp \left( - \frac{1}{2} \frac{v^2}{ \sigma_r^2} \right), \quad r \leq R_0, \label{eq:inipro1}\\
\rho_0(r) &\propto & r^n, \label{eq:inipro2}
\end{eqnarray}
with $R_0=2$ the initial radius of the sphere and the initial slope spanning the range $n=0$ to $n=-1.5$.  We consider ``warm'' and ``cool'' cases defined by their respective values of the virial ratio, $\eta=0.5$ and $0.1$, with
\begin{equation}
\eta\equiv \frac{2T}{|W|},
\end{equation}
where $T$ is the total kinetic energy and $W$ the total potential energy of the system. 

In these simulations, we aim to study in detail the evolution of the phase-space distribution function, the onset of instabilities and the consequence of these at the coarse-grained level. We shall also relate our measurements of fine details of the spiral pattern of the phase-space distribution to expectations from self-similar dynamics \citep[see, e.g.][hereafter A13]{alard13}.  To perform our simulations, we use three kinds of codes: a spherical semi-Lagrangian Vlasov solver, {\tt VlaSolve}, presented in C15, the $N$-body public treecode {\tt Gadget-2} \citep{springelg205} and a standard spherical shell $N$-body code \cite[see, e.g.][]{henon64}. Importantly, while the systems are forced to remain spherical in {\tt VlaSolve}  and in the shell code, it is not the case for {\tt Gadget-2}, which allows for the development of angular anisotropies. The variety of the codes employed in this work will help us to understand in detail the nature of different sources of instabilities, whether physical or numerical. In particular, we shall study the influence of finite spatial resolution in {\tt Vlasolve} and of finite number of particles in the shell code and in {\tt Gadget-2}.

This paper is organised as follows. In \S~\ref{sec:simus}, we present the numerical codes used to perform the simulations and provide details on the various runs we performed. In \S~\ref{sec:visu}, we perform a detailed visual inspection of the phase-space distribution function and discuss the various dynamical phases at play. Then, section \ref{sec:selfsi} deals with self-similarity: we show how the calculations of A13 can be extended to spherical systems in a very simple way, and compare theoretical predictions on the shape of the phase-space distribution function to our numerical experiments. Finally, \S~\ref{sec:conclu} summarises and discusses the results. 

\section{The simulations}
\label{sec:simus}
In spherical symmetry, Vlasov equation can be written as:
\begin{equation}
\frac{\partial f}{\partial t} + v_r \frac{\partial f}{\partial r} + \left( \frac{j^2}{r^3} - \frac{G M(<r)}{r^2} \right) 
\frac{\partial f}{\partial v_r}  =  0, \label{eq:vlasovsph}
\end{equation}
with $r$ the spherical radius, $v_r$ the radial velocity, $j$ the conserved angular momentum and $M(<r)$ the mass contained in a sphere of radius $r$. For a given value $j$, the evolution is thus driven by the interplay between the gravitational force, dominating at large radii, and the centrifugal force $j^2/r^3$, dominating at small radii. 

To solve equation (\ref{eq:vlasovsph}), we have resorted to two numerical methods. 

Firstly and mainly, we employ the spherical Vlasov solver \texttt{VlaSolve} presented in C15. This semi-Lagrangian code is similar to that of \citet{Fujiwara1983} and thus uses the splitting algorithm of \citet{cheng76} to compute the evolution of the phase-space density on a mesh. The phase-space is divided into three-dimensional cells along radius, radial velocity, and angular momentum. A logarithmic scaling is used for radius to properly resolve the dynamics of the collapse at low radii, making the use of a minimum radius $R_{\mathrm{min}}$ necessary. To compute accurately the dynamics at low radii, the exact time spent by matter elements inside the sphere of radius $R_{\mathrm{min}}$ is computed assuming that gravitational force is negligible, which is an improvement over previous implementations which used the reflecting sphere method \citep[see, e.g.][]{Gott76,Fujiwara1983}. More algorithmic details and tests of the code can be found in C15. Unless specified later, the grid for all the simulations is such that $(N_r, N_v, N_j)=(2048,2048,128)$, where $N_r$ is the number of vertices of radius in log scale, $N_v$ is the number of vertices of radial velocity in linear scale, and $N_j$ is the number of slices of angular momentum such that the $k^{\rm th}$ slice contains fluid of angular momentum $ j_{\rm max} \left( (k-1/2)/N_j \right) ^2 $, corresponding formally to the interval $[ j_{\rm max} \left( (k-1)/N_j \right) ^2, j_{\rm max} \left( k/N_j \right) ^2]$. The computation domain is $\log_{10} R_{\rm min}\equiv -2 < \log_{10} (r) < 1.4$, $ -v_l <  v_r < v_l$ and $0 < j < 1.6$, where $v_l=2$ for the runs with a virial ratio $\eta=0.5$ and $v_l=3$ for the runs with a virial ratio $\eta=0.1$. The limits are chosen such that almost all the mass of the system is contained in the computing domain during the simulated time, except for matter elements passing inside the sphere of radius $R_{\rm min}$ or those escaping from the system at large radius. For all the simulations, the time-step was chosen to be constant, equal to $\Delta t=0.005$, a value larger than in C15 to avoid excessive diffusion due to over-frequent re-samplings of the phase-space distribution function, but we checked it is still on the safe side. 

To avoid excessive aliasing effects in the \texttt{VlaSolve} runs, we apodize the initial profile given by equation (\ref{eq:inipro2}) as follows,
\begin{eqnarray}
f(\mathbf{r},\mathbf{v}) &=& \frac{\rho_0 (r)}{(2\pi \sigma_r^2)^{3/2}} \exp \left( - \frac{1}{2} \frac{v^2}{ \sigma_r^2} \right)  \nonumber \\
& & \times \frac{1}{2} \left[ 1+ {\rm erf}\left( \frac{R_0-r}{\Delta} \right) \right],
\quad r \leq R_0, \label{eq:inipromod}
\end{eqnarray}
with $\Delta=1/2$, exactly as in C15. This apodization slightly changes the value of the virial ratio, of the order of 10 percent at most. 

The second code we use relies on the standard spherical shells approach as in e.g. \citet{henon64} and \citet{Gott76}. Note that we employ it only for the most critical cases, namely $(\eta,n)=(0.1,-1)$ and $(\eta,n)=(0.1,-1.5)$, when the results from {\tt Gadget-2} differ too much from {\tt VlaSolve}. In this $N$-body code, each particle represents a shell in configuration space and interacts with the other particles through gravitational force, $-G M(<r)/r^2$, which can be obtained very easily with a sorting procedure. The resolution of the Lagrangian equations of motion of the particles is performed simply with a leapfrog integrator with a constant time-step $dt=0.001$, five times smaller than the one chosen for the Vlasov code. Similarly as in {\tt Vlasolve}, the Leapfrog algorithm is implemented using a decomposition of the Hamiltonian of the motion into a fully analytical drift part including centrifugal force and a kick part including solely gravitational force \citep[see, e.g.][]{coltoum08}. Initial shells distribution simply consists in taking the initial conditions of the {\tt Gadget-2} simulations described below, with the radius of each shell being equal to the magnitude of the position of each particle and their respective radial velocities and angular momenta directly derived from the three coordinates of the particle velocities. 

Finally, we perform simulations using the public three-dimensional $N$-body treecode {\tt Gadget-2} \citep{springelg205} in its non cosmological set-up and with the treecode part only. The positions and velocities of particles are generated in a random way using a standard rejection method. 

In the {\tt Gadget-2} simulations, spherical symmetry is no longer imposed, which leaves room for the development of angular anisotropies due to Poisson fluctuations in the initial particle distribution, even though the profile is initially spherical in the statistical sense. In particular, we shall see that the $(\eta,n)=(0.1,-1)$ and $(\eta,n)=(0.1,-1.5)$ simulations are subject to radial orbit instability. All the simulations in this work involved 10 million particles, except that we performed an additional one with 100 million particles to examine more closely the case $(\eta,n)=(0.1,-1.5)$. The parameters for the {\tt Gadget-2} runs, in terms of softening length, force accuracy and time-step control, are otherwise the same as in C15. 

Table \ref{tab:simusdata} summarises the parameters used to perform the simulations of this work. In particular, the last two columns give the value of $25\, t_{\rm dyn}$ where
\begin{equation}
t_{\rm dyn}=\sqrt{\frac{3 \pi}{\rho}}
\end{equation}
corresponds to the duration of a full radial orbit in a harmonic potential corresponding to a fixed density $\rho$. We estimate $\rho$ from the simulations themselves, once the system has reached a quasi-stationary state, either directly at the centre of the system or as the average of the density in a sphere containing 10 percent of the total mass. Our estimates are rather crude but justify the choice of final time equal to $80$ and $40$ for the ``warm'' and ``cool'' cases, respectively. Yet, we have to stay aware of the fact that increasing the magnitude of the slope $|n|$ decreases the value of $t_{\rm dyn}$. 
\begin{table*}
\begin{tabular}{llllll}
\hline
$n$                        &      $\eta$              &  $(N_r, N_v, N_j)$             & $N$ (\texttt{Gadget-2})    &  25$\,t_{\rm dyn}$ (centre)  & 25$\,t_{\rm dyn}$ (10 percent)  \\
\hline
0                            &      0.545                    & (2048, 2048,128)             & $10^7$                         & 108                                & 123    \\
-0.5                       &       0.536                      &  ''                                     & "                                  & 82                                   & 106    \\
-1.0                        &       0.526                       &  ''                                    & "                                  & 47                                  & 77       \\
-1.5                       &       0.515                     &  ''                                     & "                                  & 13                                   & 46      \\
0                            &      0.109                    &  ''                                    & ''                                  &   17                                 & 21       \\
-0.5                       &      0.108                      &  ''                                     & ''                                 &   15                                 & 24       \\
-1.0                       &      0.106                       & ''                                      & ''     + shells              &   9                                 & 23       \\
-1.5                       &      0.104                     &  ''  + (1024,512,512)        & ''   +$10^8$ + shells     &   4                                 & 19       \\
\hline
\end{tabular}
\caption[]{Parameters used for the simulations performed in this article. From left to right, the table gives the slope $n$ of the initial density profile, the actual value $\eta$ of the virial ratio after apodization (see equation~\ref{eq:inipromod}), the resolution $(N_r,N_v,N_j)$ of the Vlasov code grid, the number $N$ of particles in the \texttt{Gadget-2} simulations, with the mention of when the shell code is used as well. Finally, the two last columns give the approximate value of time after 25 harmonic orbits respectively computed from the density measured at the centre of the simulation and from the average density measured in a sphere containing 10 percent of the total mass of the system, once the system has reached a quasi-stationary regime.}
\label{tab:simusdata}
\end{table*}

\section{Visual inspection: phase-space structure and density profiles}
\label{sec:visu}
\begin{figure*}[htp]
\centering
\includegraphics[width=17cm]{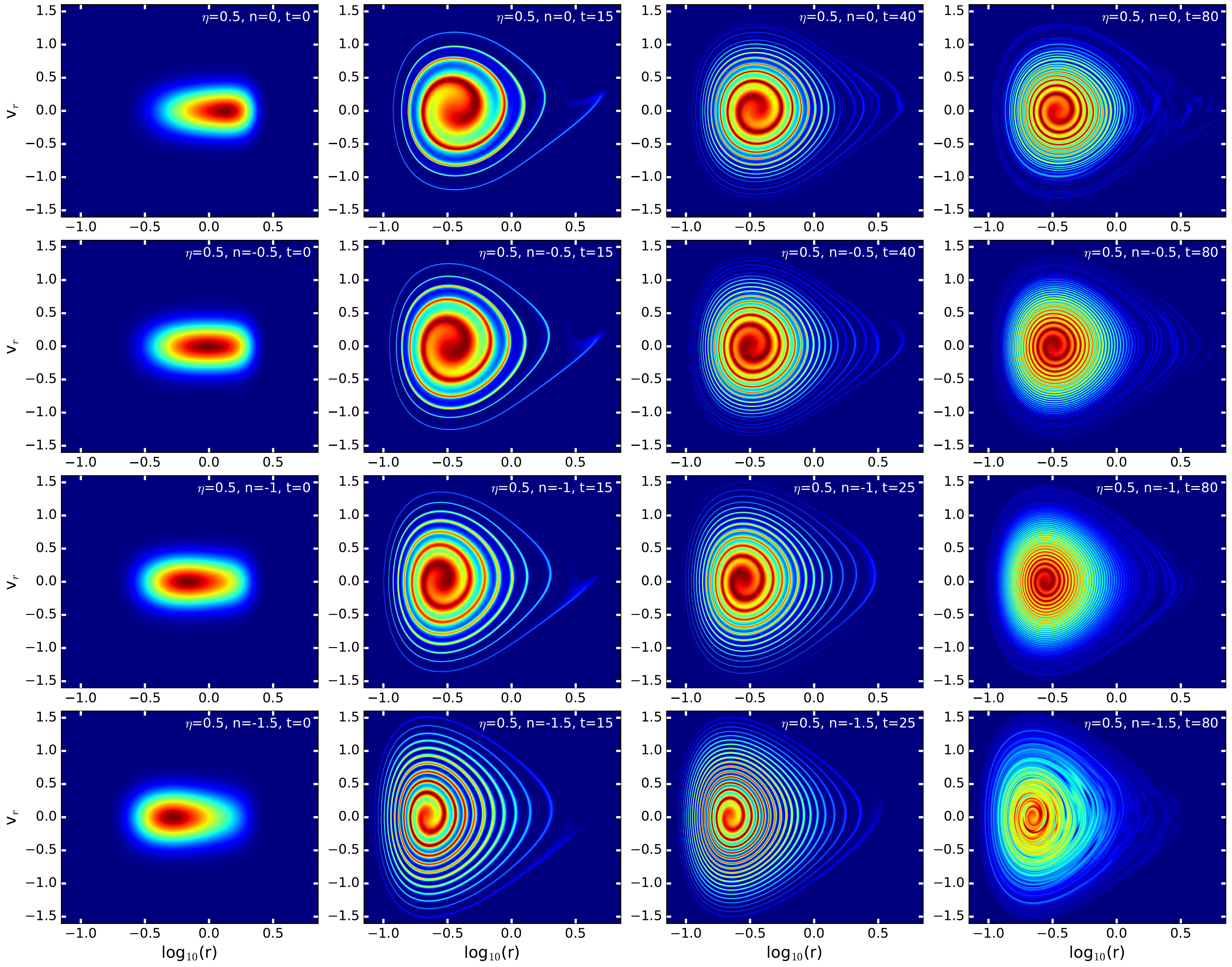}
\caption[]{Snapshots of the phase-space density for the simulations with ``warm'' initial conditions, $\eta \simeq 0.5$. A typical slice of $f(r,v_r,j)$ with $j=0.16$ is shown in $(r,v_r)$ space for the \texttt{VlaSolve} simulations at $t=0$, at an early time $t=15$, at an intermediate time used to perform tests of self-similarity of the phase-space spiral and at the final time.}
\label{fig:phase-space-warm}
\end{figure*}
\begin{figure*}[htp]
\centering
\includegraphics[width=17cm]{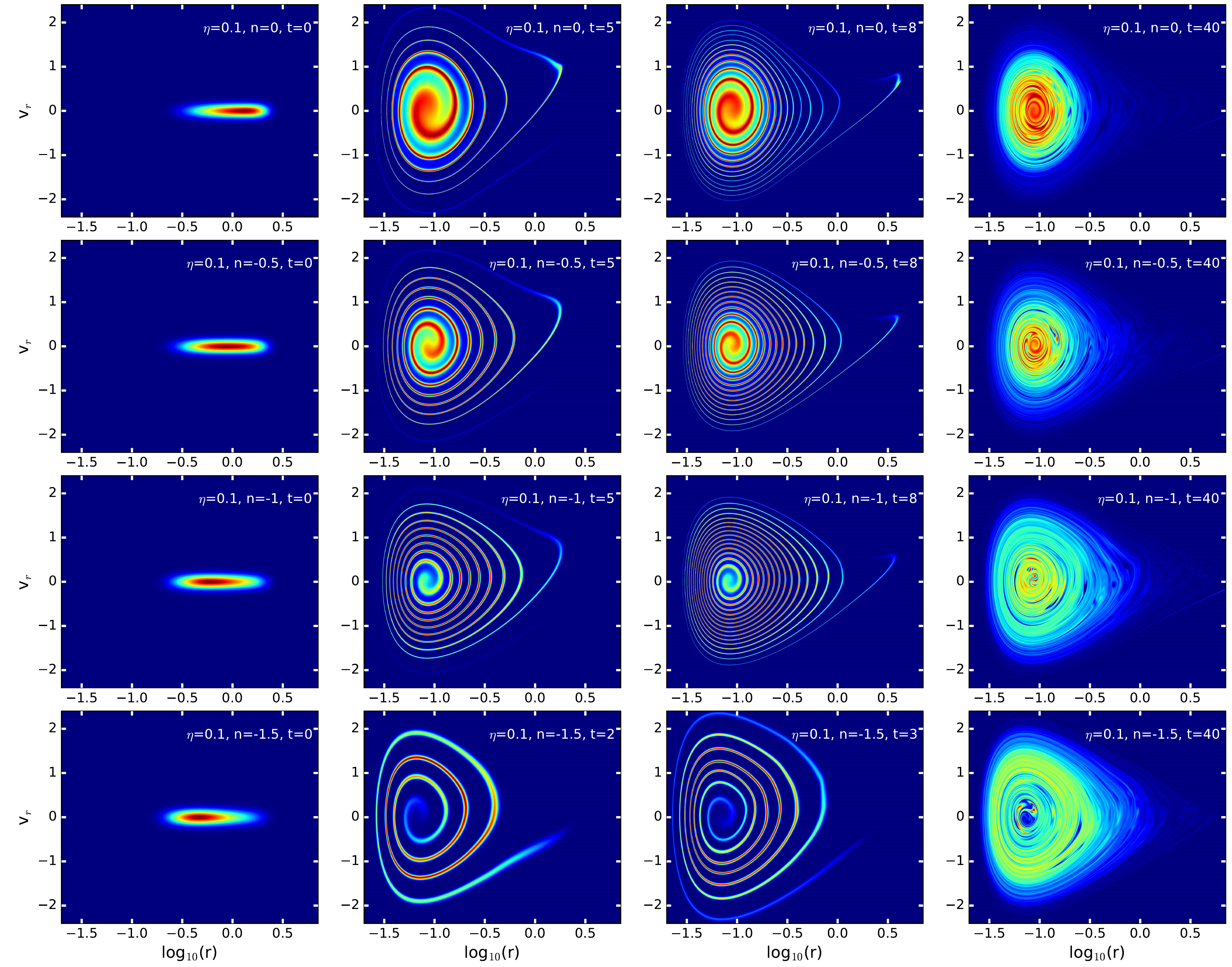}
\caption[]{Same as in Fig.~\ref{fig:phase-space-warm} but for the simulations with ``cool'' initial conditions, $\eta \simeq 0.1$ and for a slice with $j=0.06$.}
\label{fig:phase-space-cold}
\end{figure*}
\begin{figure*}[tbp]
\centering
\includegraphics[width=14cm]{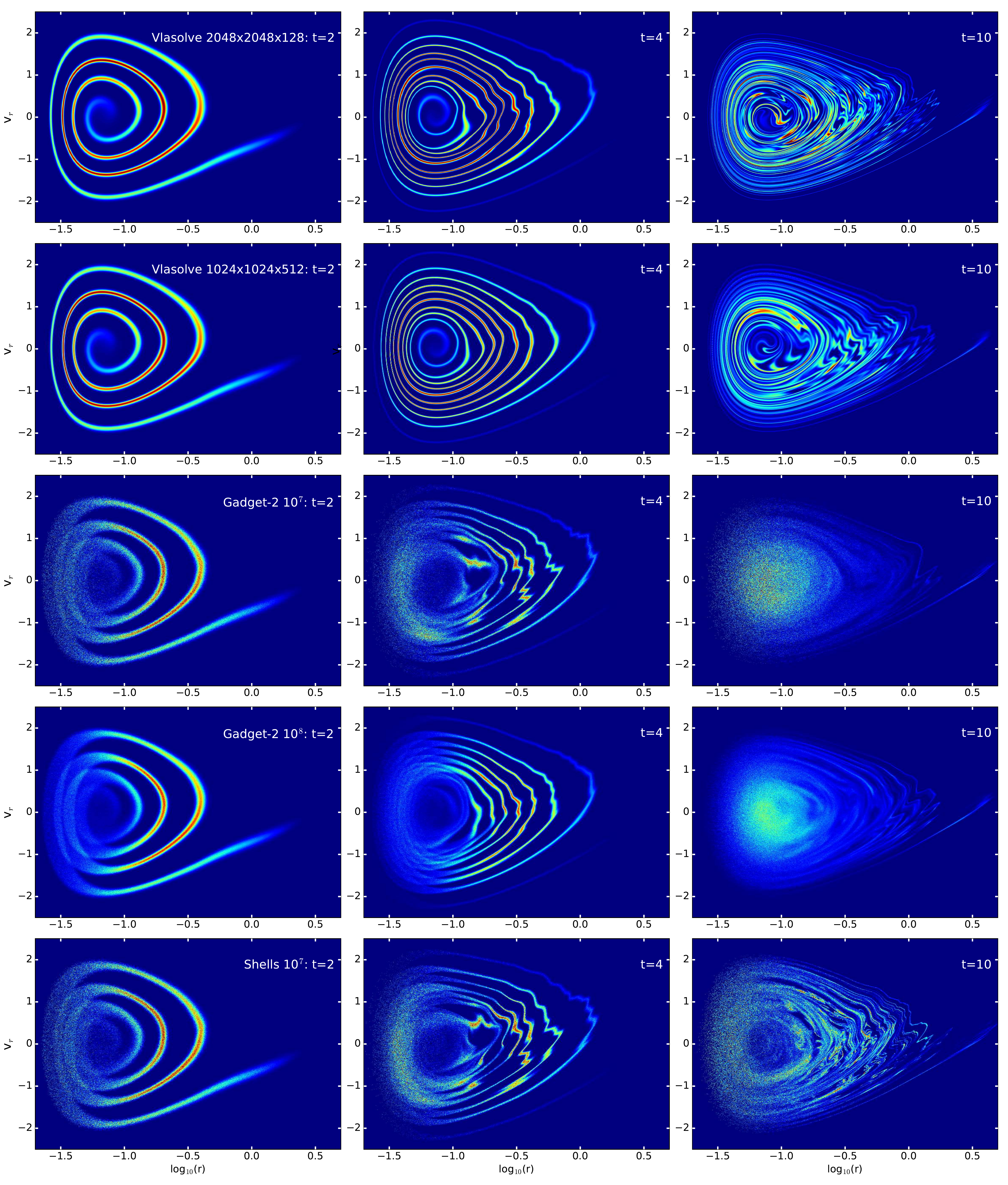}
\caption[]{Closer examination of the onset of instabilities in phase space for the $(\eta,n)=(0.1,-1.5)$ simulations: effects of spatial and mass resolutions. For the same angular momentum slice, $j=0.06$, as in lower panels of Fig.~\ref{fig:phase-space-cold}, the phase space density is represented in $(r,v_r)$ space.  The two first lines of panels correspond to two \texttt{VlaSolve} simulations with respective resolutions $(N_r,N_v,N_j)=(2048,2048,128)$ and $(1024,512,512)$. The next two lines of panels correspond to two \texttt{Gadget-2} simulations with respective numbers of particles $N=10^7$ and $N=10^8$ and the last line of panels gives, for $N=10^7$, the result obtained for the shell code. For the $N$-body simulations, the phase-space density is sampled on grids with resolution $(N_r,N_v,N_j)=(1024,1024,32)$. One can notice that the phase-space sheet is fuzzier in the $N$-body simulations than in the Vlasov code at low radius, this is because what is actually plotted is the distribution of particles (or shells) in a relatively large interval of angular momentum $j \in [0.056,0.077]$  to have sufficient number of particles to trace the phase-space distribution function, while for the Vlasov simulation, we just selected the slice corresponding to the value of $j$ of interest. This figure illustrates the effect of radial instabilities and their dependence on spatial resolution (for the Vlasov code) or mass resolution (for the $N$-body code). Note the nice agreement between the shell simulation and the {\tt Gadget-2} runs with 10 million particles in the middle panels of third and fifth lines, while they diverge in the right panels, when radial orbit instability effects become prominent in {\tt Gadget-2}.}
\label{fig:phase-space-n1v5}
\end{figure*}
\begin{figure*}[htp]
\centering
\includegraphics[width=15cm]{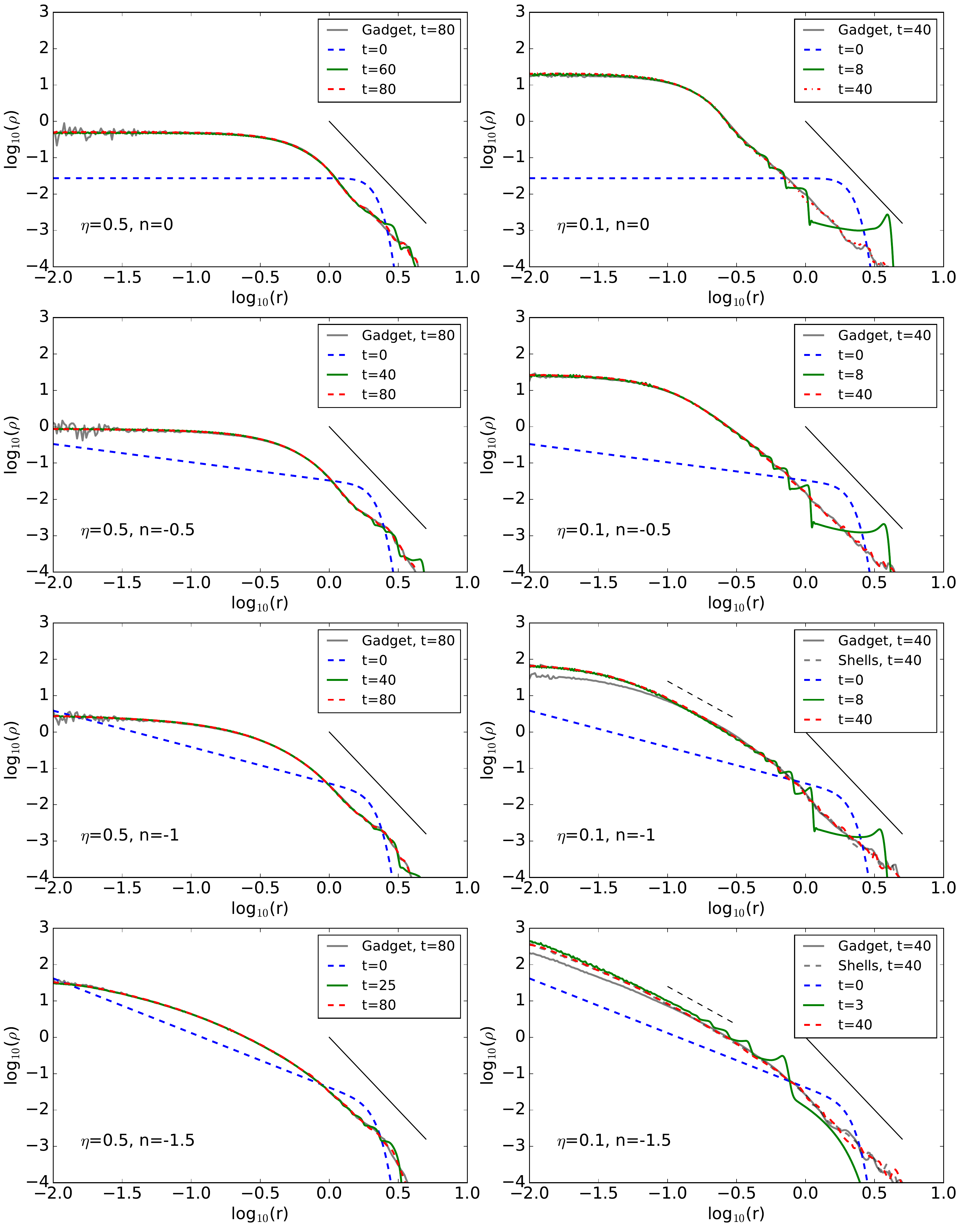}
\caption[]{Radial density profile measured at various times in the \texttt{VlaSolve} simulations, namely initial conditions (blue dashes), intermediate time used to perform tests of self-similarity of the phase-space spiral (green) and final time (red dashes). For the final time, the results are also compared to a \texttt{Gadget-2} run (thick grey), as well as the output of the shell code (thick grey dashes) for $(\eta,n)=(0.1,-1)$ and $(0.1,-1.5)$. In addition, the logarithmic slopes $-4$ and $-2.1$
\citep[as measured in][]{hozumi00} are shown respectively as a thin solid and a thin dashed line.}
\label{fig:radial_density}
\end{figure*}
\begin{figure*}[htp]
\centering
\includegraphics[width=0.99\columnwidth]{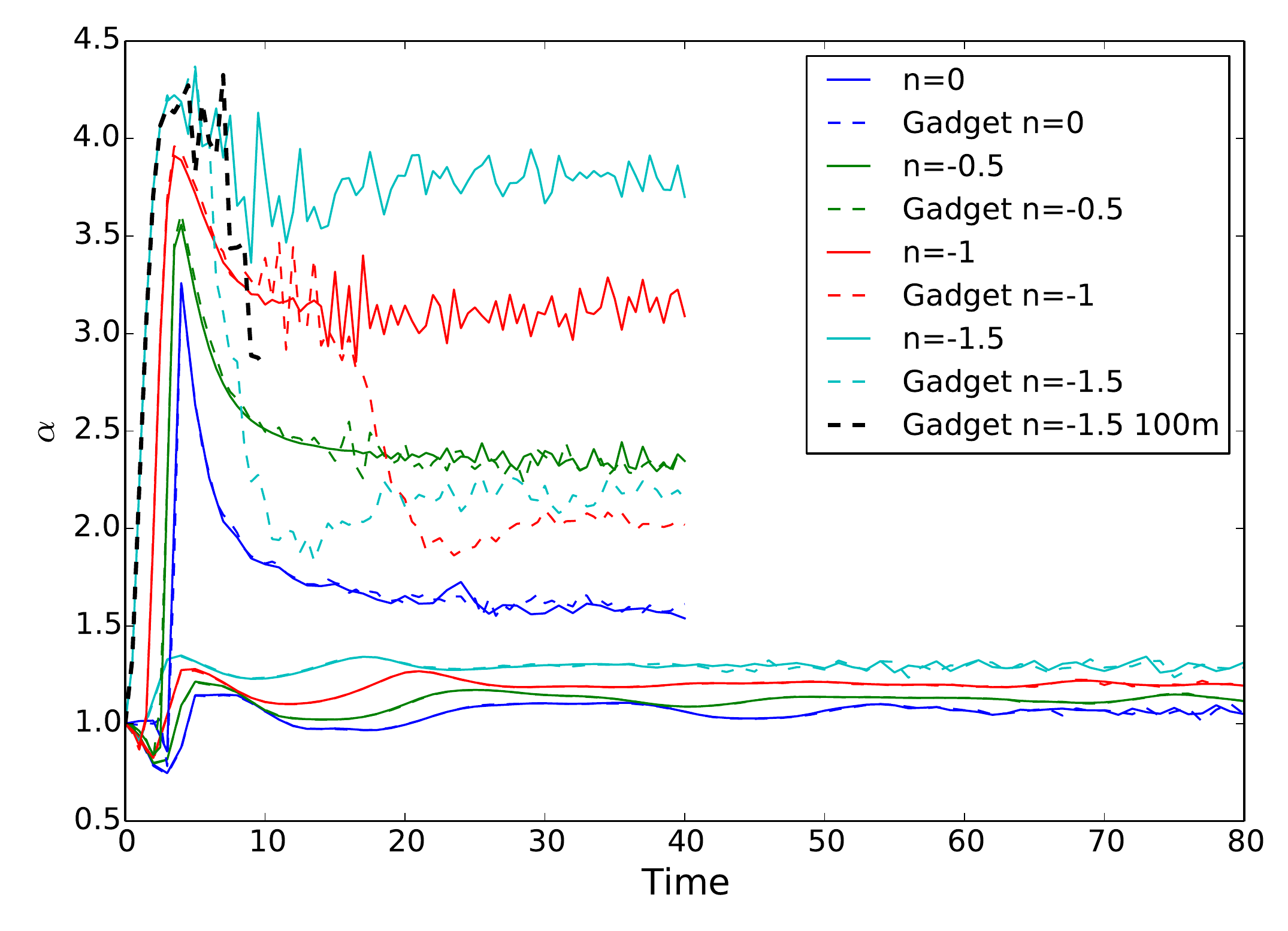}
\includegraphics[width=0.99\columnwidth]{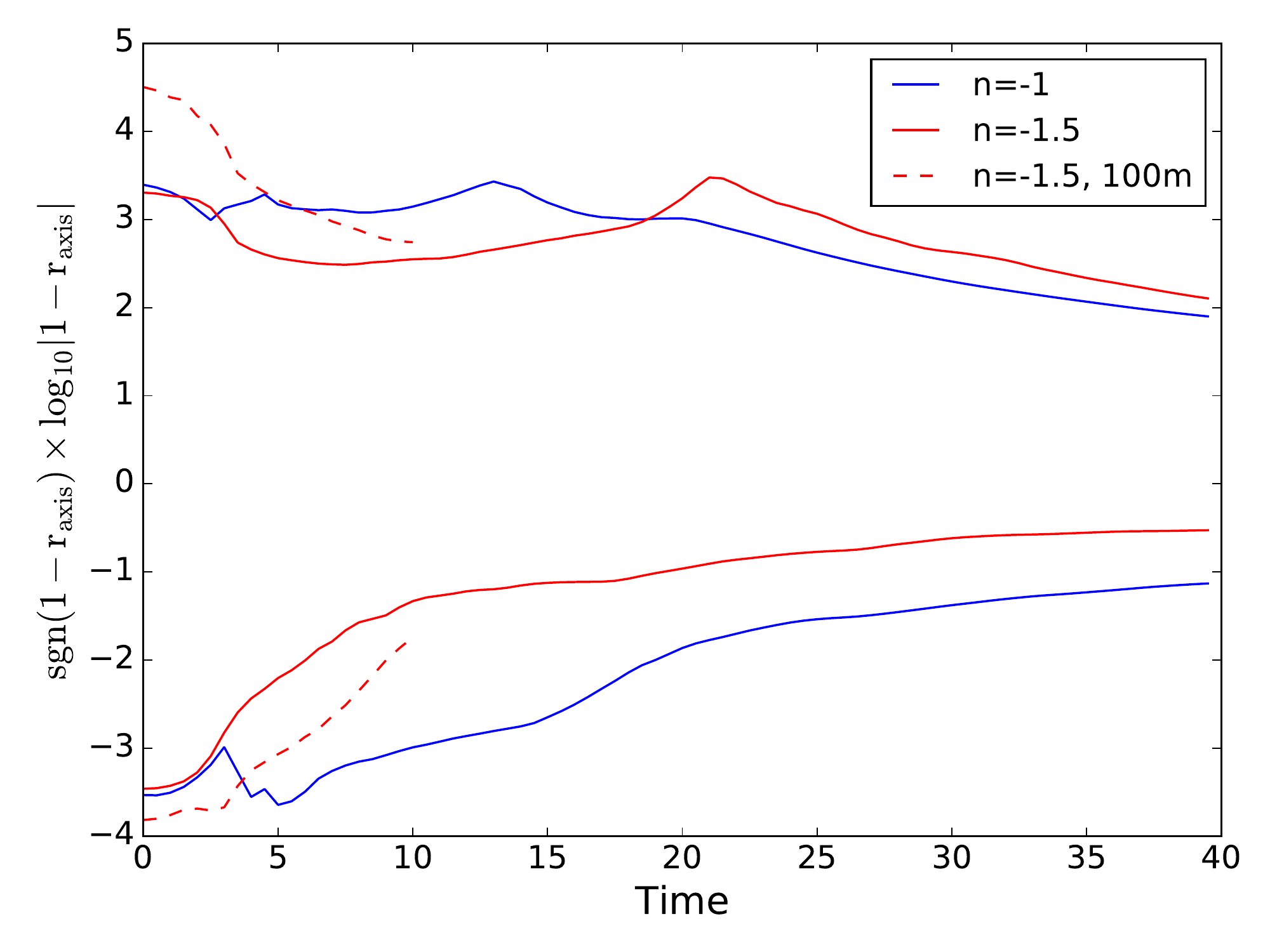}
\caption[]{Velocity anisotropy and deviation from sphericity. Left panel: velocity anisotropy parameter $\alpha=2\langle v_r^2 \rangle/\langle v_{\perp}^2 \rangle$ as a function of time for the \texttt{VlaSolve} (solid lines) and \texttt{Gadget-2} simulations we performed (dashed curves). Right panel: evolution of the departure from spherical symmetry for the two kind of initial conditions experiencing radial orbit instability in \texttt{Gadget-2}, namely $(\eta,n)=(0.1,-1.0)$ (blue curves) and $(\eta,n)=(0.1,-1.5)$ (red curves, solid and dashed for the 10 and 100 million particles simulations, respectively). The quantity $\rm{sgn}(1-r_{\rm axis}) \log_{10}|1-r_{\rm axis}|$ indicates the deviation from unity of $r_{\rm axis}$ (in log scale and with a negative sign for $r_{\rm axis} > 1 $), where $r_{\rm axis}=b/c$ (upper curves) or $b/a$ (lower curves), and $a \leq b \leq c$ are the principal axis lengths of the \texttt{Gadget-2} particle distribution derived from the inertia tensor.} 
\label{fig:anisotropy}
\end{figure*}
Figures \ref{fig:phase-space-warm} and \ref{fig:phase-space-cold} display, for a typical slice of fixed angular momentum, the phase-space distribution function of the ``warm'' and ``cool'' \texttt{VlaSolve} simulations, respectively.  Additionally, Figure~\ref{fig:phase-space-n1v5} examines more in detail the case $(\eta,n)=(0.1,-1.5)$, which is subject to radial orbit instability (hereafter ROI), while Figure~\ref{fig:radial_density} provides projected density profiles. To supplement our discussion about ROI, we study in Fig.~\ref{fig:anisotropy} the velocity anisotropy parameter for all the simulations as well as deviation from sphericity for the runs which experience ROI. 

Thanks to the high resolution of our simulations, when examining these figures, one can clearly separate, for the first time, 2 or 3 well known dynamical phases, depending on initial conditions: (i) a violent relaxation phase during which the system converges to a quasi-steady state by building a very regular spiral structure in phase-space, (ii) a quiescent phase during which the quasi-steady state is preserved against small scale radial instabilities which can destroy the spiral and (iii) relaxation to a non spherical state through ROI when the system is prone to develop it. The novelty in our measurements is obviously not the discovery of the various phases of the dynamics, which are heavily discussed in the literature, but instead the clear articulation between them for the systems we study.  We discuss now these three phases in detail.  
\subsection{Violent relaxation} 
In a first phase, the system undergoes violent relaxation that leads quickly to the establishment of a quasi-steady state. During this phase, spherical symmetry is preserved and the phase-space distribution function presents in all the cases a very regular spiral structure, even in the $N$-body runs, thanks to the large number of particles we used to perform them. A visualisation of a film of the evolution of the system shows that it is subject as well to a global pulsation that introduces at some point irregular features in the phase-space distribution function in the outer parts of the spiral and at large radius, e.g. the dark region in upper right panel of Fig.~\ref{fig:phase-space-warm}. During this violent relaxation phase, \texttt{Gadget-2} agrees very well with \texttt{VlaSolve}, even in regions where these irregular features develop, as already noticed by C15 for the $(\eta,n)=(0.5,0)$ case, which shows that these features are intrinsic to the physical system and are not related to some additional instability due to some numerical noise.  
\subsection{Quasi-steady regime with small scale radial instabilities} 
In a second phase, the system stays in quasi-equilibrium and preserves its spherical symmetry. However, some radial instabilities perturb it at small scales, whatever numerical technique used, and can destroy the spiral structure. The time of the appearance of these instabilities is related to spatial resolution in the {\tt Vlasolve} simulation and to the number of particles in the $N$-body simulations. This is well illustrated by Fig.~\ref{fig:phase-space-n1v5} for $(\eta,n)=(0.1,-1.5)$: for instance, increasing the number of angular momentum slices in the \texttt{VlaSolve} simulation reduces the magnitude of the perturbations of the spiral (compare middle insert of first and second line of panels), and similarly when increasing the number $N$ of particles in the \texttt{Gadget-2} runs (compare middle insert of third and fourth line of panels). 

In the $N$-body case, these collective instabilities are induced by small random but correlated errors on the gravitational force due to Poisson fluctuations in the particle distribution. In the {\tt Vlasolve} code, they are related to coherent errors on the force due to the representation of the phase-space density on a grid, but the effect is analogous to the $N$-body case. These instabilities become naturally more significant when the initial velocity dispersion is reduced, as explained in C15. We also notice here that they take place earlier for larger $|n|$, in agreement with our calculation of dynamical times in the two right columns of Table~\ref{tab:simusdata}. As a result, during the interval of time we run our simulations, they can be seen in the cool runs and in the warm case for $(\eta,n)=(0.5,-1.5)$, but they are not present or negligible in other cases. 

The important fact is that these instabilities intervene only at the fine level: they do not change the structure of the system at the coarse level, even quantitatively. To be more specific, if the phase-space density was smoothed at scales larger than the inter-filament separation --by filament, we mean e.g. some fold of the spiral structure-- and than the size of the fluctuations introduced by radial instabilities, there would be no significant difference between late times, where these instabilities can destroy the spiral structure, and earlier times, when the spiral structure is still well defined. A good way to illustrate this consists in examining the projected density profile measured in the {\tt Vlasolve} simulations, as displayed in Fig.~\ref{fig:radial_density}, and to compare red curves to the green ones, that correspond respectively to these aforementioned late and earlier times. The calculation of the projected density, by integrating the phase-space density over velocities, indeed corresponds to some coarse-graining procedure, although such anisotropic smoothing does not erase the quasi-caustic structures seen on the green curves of Fig.~\ref{fig:radial_density}.  These bumps correspond to projection of parts of the spiral (or any filament) that are orthogonal to configuration space. However, with proper (adaptive) smoothing at scales larger than the space between successive spiral folds, one can be convinced that agreement between the green curves and the red curves, already very good in most cases, should improve furthermore.

Hence, the quasi-equilibrium built cinematically by the spiral is stable against radial perturbations, but not necessarily the spiral structure.
\subsection{Deviation from sphericity: radial orbit instability} 
In a third phase, small angular anisotropies induced by numerical noise get amplified through radial orbit instability (ROI) for  $n \leq -1$ in the cool cases and the system deviates from spherical symmetry by acquiring a prolate shape (right panel of Fig.~\ref{fig:anisotropy}).  A consequence of ROI is the reduction of the magnitude of the spherically averaged density profile $\rho(r)$ at small radius, as can be seen on two bottom right panels of Fig.~\ref{fig:radial_density}.  

ROI signature is best seen in the velocity anisotropy parameter
\begin{equation}
\alpha=\frac{2\langle v_{r}^2 \rangle}{\langle v_{\perp}^2 \rangle}
\end{equation}  
\citep[see, e.g.][]{Hozumi96}, where $v_r$ and $v_{\perp}$ are respectively the radial and transverse velocities, as plotted in left panel of Fig.~\ref{fig:anisotropy}. 
Due the dominant nature of radial infall during the very first phase of violent relaxation, cool initial conditions induce, after collapse, a strong velocity anisotropy, which is known, when exceeding some (still not fully known) threshold, to trigger ROIs in presence of small perturbations to spherical symmetry \citep[see, e.g.][and references therein]{Polyachenko1981,Merritt1985,Barnes1986,Barnes2009,Marechal2011,Polyachenko2015}. In this case, the onset of ROI reduces significantly the value of $\alpha$, as illustrated by left panel of Fig.~\ref{fig:anisotropy} for $(\eta,n)=(0.1,-1)$ and $(\eta,n)=(0.1,-1.5)$.

In our {\tt Gadget-2} simulations, small perturbations from spherical symmetry are related to shot noise, so the onset of ROI is particle number dependent \cite[see, e.g.][]{Benhaiem2018}, as illustrated on Fig.~\ref{fig:anisotropy} by our two runs with 10 and 100 million particles in the $(\eta,n)=(0.1,-1.5)$ case. On this figure, one also notices that ROI takes place later for $n=-1$ than for $n=-1.5$, but this is roughly consistent with the dynamical times given in the last two column of Table~\ref{tab:simusdata}. 

The conditions of establishment of radial orbit instability are however not yet fully understood: some theoretical calculation and numerical experiments show that it should take place when $\alpha > \alpha_{\rm critical}$ with $\alpha_{\rm critical}$ ranging between $1$ and $2.9$ \citep[see, e.g.][]{Polyachenko2015}: this condition is clearly satisfied for $(\eta,n)=(0.1,-1.0)$ and $(0.1,-1.5)$ when examining left panel of Fig.~\ref{fig:anisotropy}. Strictly speaking, given the limited amount of time we run the simulations, the other cases remain undecided even though we do not detect any ROI. The results obtained elsewhere in the literature, in particular by \cite{Merritt1985}, \cite{Barnes2009}, suggest that our ``warm'' systems are probably not prone to ROI, while, for $(\eta,n)=(0.1,-0.5)$ and $(\eta,n)=(0.1,0)$, there is still a chance that ROI develops after some time. Clearly, our simulations are not run long enough to have all the details of the history of the system, which might evolve further to another interesting state. 

Finally, note that whether pure radial instability takes place before ROI is difficult to quantify in our simulations.  Using linear analysis during collapse phase, \citet[][]{Aarseth1988} argue that in the cold case, angular anisotropies introduced by Poisson noise are sub-dominant compared to radial ones, which suggests that a radial instability phase could take place before ROI. This argument is partly supported by Fig.~\ref{fig:phase-space-n1v5}, where excellent agreement is found between {\tt Gadget-2} and the spherical shell code at $t=4$ (middle inserts of 3rd and 5th lines of panels), when radial perturbations already significantly disrupt the phase-space spiral while ROI did not develop yet. 

\section{Self-similarity in phase-space}
\label{sec:selfsi}

In practice, seeking self-similar solutions to Vlasov-Poisson equations consists in finding solutions invariant with respect to some homothetic transforms, e.g., 
\begin{equation}
f(\lambda_1 \mathbf{x}, \lambda_2 \mathbf{v}, \lambda_3 t)=\lambda_4 f(\mathbf{x},\mathbf{v},t),
\label{eq:selfs}
\end{equation}
which requires in this case $f$ to be of the form of 
\begin{equation}
f(\mathbf{x},\mathbf{v},t)=t^{\alpha_0} F\left( \frac{\mathbf{x}}{t^{\alpha_1}}, \frac{\mathbf{v}}{t^{\alpha_2}} \right).
\end{equation}
Solving Vlasov equation provides the solution for function $F$. While a rigorous framework can be set to define self-similarity through Lie derivatives \citep{Carter1991}, there is no unique way to express it. For instance, as studied in e.g. \cite{Henriksen1995}, instead of equation (\ref{eq:selfs}), one can, in the spherically symmetric case, introduce anisotropy in the self-similar solution by separating the radial velocity from the angular momentum, or, in an extreme but well known case, one can just assume, as in the cold case, pure radial motions with zero angular momentum.

Strictly speaking, self-similarity implies a pure power-law behaviour for the projected density \citep[see, e.g.][]{Henriksen1995}, which is obviously not the case for the systems we study here when examining Fig.~\ref{fig:radial_density}, except to some extent at sufficiently large radii. However, self-similarity usually takes place only in a limited domain of phase-space. This can for instance simply be due to the finite extension of the system, which may induce deviations of the density profile from a pure power-law, even though phase-space density is perfectly self-similar. As a clear illustration of this state of fact, \citet[]{alard13} argues that despite the cut-offs due to the finiteness of the system, local self-similarity in phase-space implies a power-law behaviour for the quantity $Q(r)=\rho(r)/\sigma(r)^3$ where $\rho(r)$ and $\sigma(r)$ are respectively the projected density and the velocity dispersion, even if these latter are not found to be exact power-laws of radius due to the cut-offs.  Such a power-law property for $Q(r)$ is verified to a great accuracy by dark matter halos \citep[see, e.g.][]{Taylor2001}, of which the density profiles are known to deviate from pure power-laws \citep[]{NFW1,NFW2}. As shown by \citet{Dehnen2005}, solving the spherical Jeans equation assuming that $Q(r)$ is a power-law can indeed lead to non pure power-law density profiles. 

\begin{figure*}[htp]
\centering
\includegraphics[width=0.99\columnwidth]{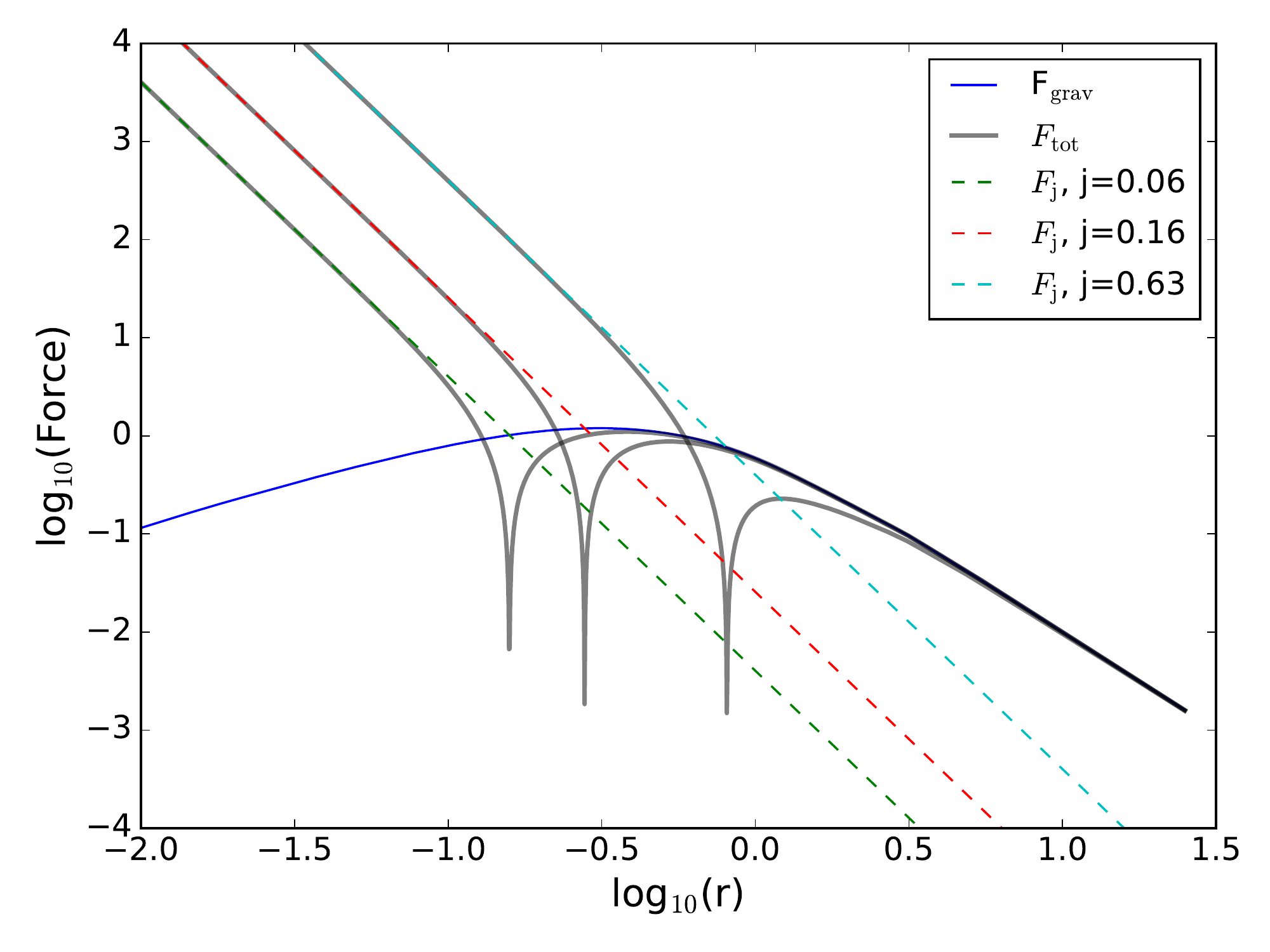}
\includegraphics[width=0.99\columnwidth]{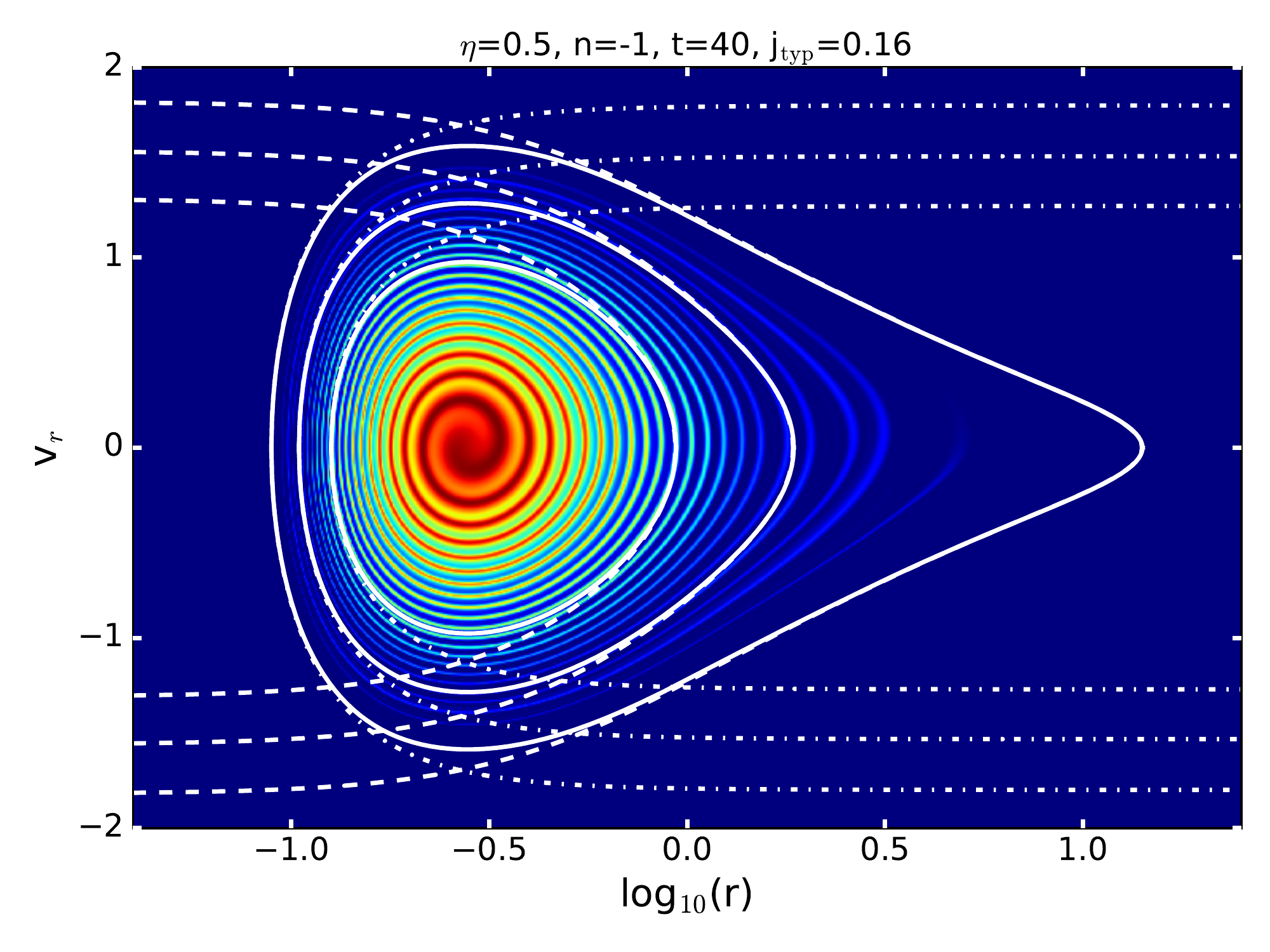}
\caption[]{The small and large radii regimes in the $(\eta,n)=(0.5,-1)$ \texttt{VlaSolve} run at $t=40$. Left panel plots separately the magnitude of  centrifugal force $j^2/r^3$ and of the gravitational force as functions of radius as well as the magnitude of the sum of both forces. The right panel shows isocontours of the specific energy (white curves) superposed on the phase-space distribution function for $j=0.16$. In addition, the white dashes and dot-dashes assume respectively that only the gravitational or the centrifugal force contributes.}
\label{fig:regimesJG}
\end{figure*}

Here, we are clearly not in the case of a single self-similar regime (with possible cut-off effects). Indeed, for each value of the angular momentum $j$, our systems can trivially be separated into two distinct regions of phase-space. In the first region, corresponding to small enough radius, the centrifugal acceleration, $j^2/r^3$, dominates. Its power-law nature is expected to induce self-similarity in some domain of the considered phase-space slice. In the second region, corresponding to large enough radius, the gravitational acceleration, $-G M(<r)/r^2$, dominates.  Provided that it is also a power-law of radius, one expects another self-similar behavior. Given the discussion above about the possible effects due to the finite extent of the system, self similar properties may be found even if the force is not exactly a power-law. The transition between these two regions is sharp, as illustrated by Fig.~\ref{fig:regimesJG}, which is important to make our approach meaningful. 

Self-similarity also predicts the set-up of a spiral in phase-space, of which the structure is defined by the self-similar parameters \cite[see, e.g.][A13]{fillmore84}. Note, as already mentioned, that the onset of self-similarity does not need assuming cold initial conditions as it is often supposed. For instance, in the calculations of A13, no such hypothesis is made, and the existence of a spiral structure in phase-space is clearly evidenced just by assuming self-similarity.  Here, we cannot rigorously demonstrate the existence of such a spiral structure but can postulate it, predict its local properties in each of the supposed self-similar regimes mentioned above and compare the predictions to our simulation measurements, which we do now, following closely \citet[][2016]{alard13}. 

When examining a slice of fixed angular momentum, we notice that the Vlasov equation for a spherical system is exactly analogous to the one-dimensional case:
\begin{equation}
\frac{\partial f}{\partial t} + v_r \frac{\partial f}{\partial r}  -\frac{\partial {\psi}}{\partial r} \frac{\partial f}{\partial v_r}  =  0, \label{eq:vlasovsph2}
\end{equation}
except that the force derives from the following scalar field
\begin{equation}
\psi(r)=\frac{j^2}{2r^2} + \phi(r),
\end{equation}
where $\phi(r)$ is the gravitational potential. A13 derived detailed self-similar solutions in the 1D case that we extend below in the regimes where the centrifugal force dominates $\psi(r) \simeq j^2/(2 r^2)$ and in the regime where gravitational potential dominates and is a power-law, $\psi(r) \simeq \phi(r) \propto r^{\beta+2}$. 

Assuming that the conserved angular momentum $j$ is a dummy variable, the self-similar solution for the phase-space distribution function can be expressed as follows:
\begin{equation}
f(r,v_r,j, t)=t^{\alpha_0} F_j\left( \frac{r}{t^{\alpha_1}},\frac{v_r}{t^{\alpha_2}} \right). 
\end{equation}
Setting 
\begin{equation}
{\tilde r}=\frac{r}{t^{\alpha_1}}, \quad {\tilde v}=\frac{v_r}{t^{\alpha_2}},
\end{equation}
we can express both the gravitational and the centrifugal force as functions of these new variables.

Starting from the gravitational force, we define a function $U$ such that:
\begin{equation}
- \frac{\partial \phi}{\partial  r}  \equiv t^{\alpha} U({\tilde r}),
\end{equation}
and we assume that 
\begin{equation}
U=-{{\rm d} {\tilde \phi}}/{{\rm d}{\tilde r}},
\end{equation}
with
\begin{equation}
{\tilde  \phi}({\tilde r})  \equiv k\, {\tilde r}^{\beta+2}.
\label{eq:phider}
\end{equation}
Then,
\begin{equation}
U({\tilde r})=-k (\beta+2) {\tilde r}^{\beta+1},
\label{eq:upow}
\end{equation}
and we obtain from Poisson equation:
\begin{eqnarray}
U({\tilde r}) & = & - \frac{G}{{\tilde r}^2} \int_{{\tilde r}' < {\tilde r}} 8 \pi^2 F_j({\tilde r}',{\tilde v}) {\rm d}{\tilde r}' {\rm d}{\tilde v} j {\rm d} j, \\
\alpha & = & \alpha_0+\alpha_2-\alpha_1.
\end{eqnarray}

Similarly, the centrifugal force can be written
\begin{equation}
\frac{j^2}{r^3}  \equiv  t^{\alpha} U({\tilde r}),
\end{equation}
which implies
\begin{equation}
\alpha= -3\alpha_1, \label{eq:3a}
\end{equation}
and, if we assume again that $U({\tilde r})=-k (\beta+2) {\tilde r}^{\beta+1}$, we obtain:
\begin{eqnarray}
\beta&=&-4, \\
k&=&\frac{j^2}{2}.
\end{eqnarray}

By injecting these various expressions in the Vlasov equation one obtains, in a regime where either the gravitational or centrifugal force dominates,
\begin{eqnarray}
\alpha_0 F_j &+& \frac{\partial F_j}{\partial {\tilde r}} \left[ -\alpha_1\, {\tilde r} + {\tilde v}\, t^{\alpha_2-\alpha_1+1} \right] \nonumber \\
&+&\frac{\partial F_j}{\partial {\tilde v}} \left[ -\alpha_2\, {\tilde v}-k\,(\beta+2)\,{\tilde r}^{\beta+1} t^{\alpha-\alpha_2+1}\right]=0.
\label{eq:vla2}
\end{eqnarray}
Eliminating time dependence in this equation imposes
\begin{eqnarray}
\alpha_1=\alpha_2+1,\\
\alpha=\alpha_2-1.
\end{eqnarray}
Enforcing stationarity of the force $t^{\alpha} U(r/t^{\alpha_1})$ with the power-law (\ref{eq:upow}) implies
\begin{equation}
\alpha-(\beta+1)\alpha_1=0.
\end{equation}
Hence the only viable solution is
\begin{eqnarray}
\alpha_2 &=&-\frac{\beta+2}{\beta}, \label{eq:alpha2beta}\\
\alpha_1&=&-\frac{2}{\beta},\\
\alpha&=&-\frac{2\beta+2}{\beta},
\end{eqnarray}
which is of course consistent with equation (\ref{eq:3a}) and leaves $\alpha_0$ as a free parameter if the centrifugal force dominates, while, if the gravitational force dominates, it fixes $\alpha_0=-(\beta+2)/\beta$. Note that, in general, total mass is not conserved. Indeed, enforcing total mass conservation (or mass conservation per angular momentum slice, as well), imposes $\alpha_0+\alpha_1+\alpha_2=0$, a condition which is fulfilled only for $\beta=-3/2$ and in this case $\alpha_0=1/3$. This is not a problem because self-similarity is expected to take place only in a finite dynamical range.

To follow as closely as possible the notations of A13,  we now make the following change of variables,
\begin{eqnarray}
{\cal G} &\equiv& \ln F_j,\\
\eta &\equiv& \frac{\beta}{2}+1,\\
u &\equiv & {\tilde r}^\eta.
\end{eqnarray}
Equation (\ref{eq:vla2}) becomes
\begin{eqnarray}
-\alpha_0 &+&(1+\alpha_2)\, \eta\, \frac{\partial {\cal G}}{\partial u} u + \alpha_2 \frac{\partial {\cal G}}{\partial {\tilde v}} {\tilde v} \nonumber \\
               &-& \eta \left[ \frac{\partial {\cal G}}{\partial u} {\tilde v} - 2\, k\, u \frac{\partial G}{\partial {\tilde v}} \right] u^{\beta/(\beta+2)}=0,
\end{eqnarray}
which is exactly the same equation as equation 11 of A13, except that the first term $\alpha_2+2$ is replaced here with $-\alpha_0$. Hence, the solution of this equation is very similar to the expressions given in A13. The main difference here is that the values of $\beta$ we consider are outside the domain of validity of the calculations of A13, which implies that the isocontours of the solutions are closer to hyperbolic curves than to a spiral. However, here, we have to take into account the fact that we have two distinct supposed self-similar regimes, one dominated by centrifugal force, say for $r \la r_{\rm crit}(j)$, and the other dominated by gravitational force, say for $r \ga r_{\rm crit}(j)$. Hence, the actual solution is the connection between too partial solutions following self-similar properties. 
\begin{figure}[htp]
\centering
\includegraphics[width=0.99\columnwidth]{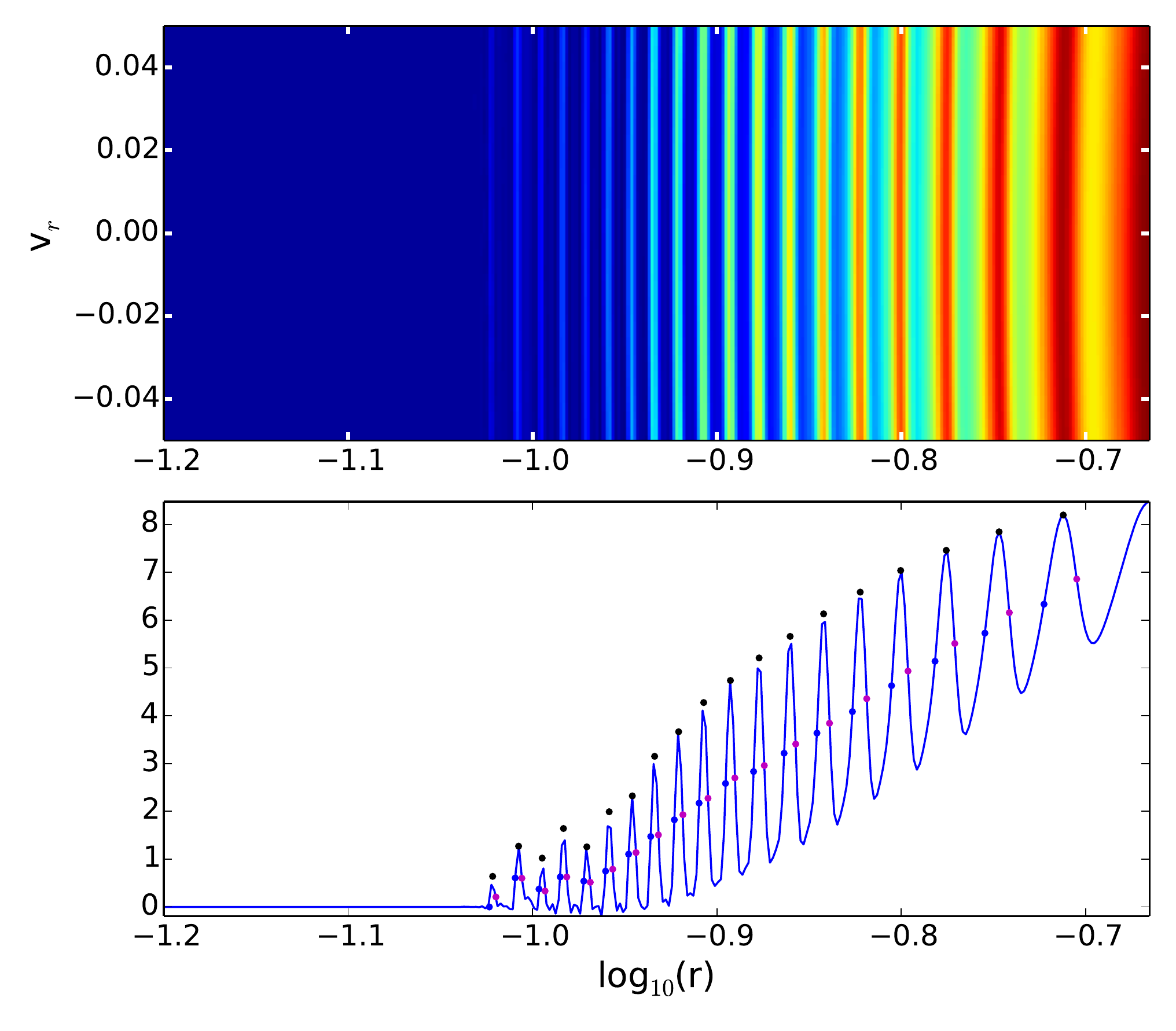}
\caption{Illustration of the method used to determine the positions of the folds and corresponding interfold distance law at small radius in a phase-space slice. On top panel, a zoom is performed around the axis $v_r=0$ in the region dominated by centrifugal force for the $(\eta,n)=(0.5,-1.0)$ simulation at $t=40$. The corresponding phase-space distribution function $f(r,v_r=0,j=0.16)$ is plotted on lower panel. The black dots give the positions of local maxima estimated with our local quadratic fit, while the blue and red dot provide upper and lower bounds to compute the (very conservative) error bars shown on Fig.~\ref{fig:fit_interfold}.}
\label{fig:vlapeaks-fig}
\end{figure}

Rescaling variable ${\tilde r}$ so that $k=1/2$ in equation (\ref{eq:phider}), and introducing, exactly as in A13, the new variables
\begin{eqnarray}
R&=&\sqrt{u^2+{\tilde v}^2},\\
\cos \Psi&=&\frac{u}{R},
\end{eqnarray}
we obtain nearly exactly equation 12 of A13, but the parameters of this equation change according to whether the value of $R \cos \Psi$ is above or below a threshold fixed by $r_{\rm crit}$. With $H(R,\Psi) \equiv G(u,{\tilde v})$ we write, following exactly the footsteps of A13, the general solution for $H$ when the power-law force is stationary,
\begin{eqnarray}
H(R,\Psi)&=&\frac{\alpha_0}{\alpha_2} \ln R  \nonumber \\
&+ & Q\left( R^{-1/\alpha_2}+\frac{1+\alpha_2}{\alpha_2} \int (\cos \Psi)^{1/\alpha_2} {\rm d}\Psi \right),
\label{eq:solH}
\end{eqnarray}
with $\Psi \in ]-\pi/2,\pi/2[$ and where $Q$ is some function. 
At this point, introducing the same concept of spiral as in A13 is not simple, because the fact that $u > 0$ does not allow $\Psi$ to make a full excursion on the circle. Furthermore, the values of the logarithmic density profile slope $\beta$ we have to consider range in the interval $-4 \la \beta < -2$, which implies, from equation (\ref{eq:alpha2beta}), $-1/2 \la \alpha_2 < 0$, hence some divergence of the integral
\begin{equation}
I \equiv \frac{1+\alpha_2}{\alpha_2} \int (\cos \Psi)^{1/\alpha_2} {\rm d}\Psi,
\end{equation}
when $|\Psi|$ approaches $\pi/2$.  This is however not a real problem, because the objective is to connect two self-similar solutions. Here, we are unable to demonstrate the existence of the spiral structure in phase-space, we have to postulate it.  We therefore define a new angular variable $\theta$ and 
\begin{equation}
I(\theta) \equiv \int g(\theta') {\rm d} \theta',
\end{equation}
where $g(\theta)$ is a function of period $2\pi$ verifying
\begin{eqnarray}
g(\theta) & \simeq & g_{-}(\theta) \equiv \frac{1+\alpha_2^{-}}{\alpha_2^{-}} [\cos (\Psi^{-}=\theta-2k \pi)]^{1/\alpha_2^{-}},  \nonumber \\
& & \quad \quad \quad \quad \quad \quad  \quad \theta-2k \pi \simeq 0,\\
g(\theta) & \simeq & g_{+}(\theta) \equiv \frac{1+\alpha_2^{+}}{\alpha_2^{+}} [-\cos (\Psi^{+}=\theta-2k \pi)]^{1/\alpha_2^{+}}, \nonumber \\
& & \quad \quad \quad \quad \quad \quad \quad \theta -2k\pi \simeq \pi,
\end{eqnarray}
and $-$ and $+$ correspond respectively to the regimes dominated by the centrifugal and the gravitational force. Function $g(\theta)$ makes a smooth transition between $g_{-}$ and $g_{+}$. The only, trivial but important fact we have to know, is that $I(\theta)$ defined this way is roughly proportional to $\theta$ which allows us now to define explicitly the concept of a spiral across both self-similar domains. The interesting bit is that the subsequent calculations of A13 are not changed at all when taking this new definition of $I$ and his equation 19 still stands in each self-similar domain, with $I_1 = \int_0^{2\pi} g(\theta') {\rm d} \theta'$ now an unknown constant instead of a well defined integral as in A13. 

Hence, we have, in the situation where there are many folds, the following expected relationship for the interfold distance in each self-similar region:
\begin{equation}
{\rm d} R \propto R^{1+1/\alpha_2}.
\label{eq:interf}
\end{equation}
In particular, coming back to standard variables $(r,v_r)$, the interfold distances along the axis $v_r=0$ reads
\begin{equation}
{\rm d}r \propto r^{1-\beta/2}.
\end{equation}
To test this property directly, we analyse, at a time where the spiral structure is still well defined, function $f(r,v_r=0,j)$ for a fixed value of angular momentum, as illustrated by Fig.~\ref{fig:vlapeaks-fig}. We determine the positions of the folds using local parabolic fits. For each fold $i$, we determine two semi-heights radial positions $\log (r_{{\rm l},i})$ and $\log (r_{{\rm r},i})$ (the computational grid being logarithmic in radius), on the left and on the right of the peak (respectively), and define the error on the position of the peak as $\delta \log (r_i) = \log (r_{{\rm r},i}) - \log(r_{{\rm l},i}) $. 

\begin{figure*}[htp]
\centering
\includegraphics[width=17cm]{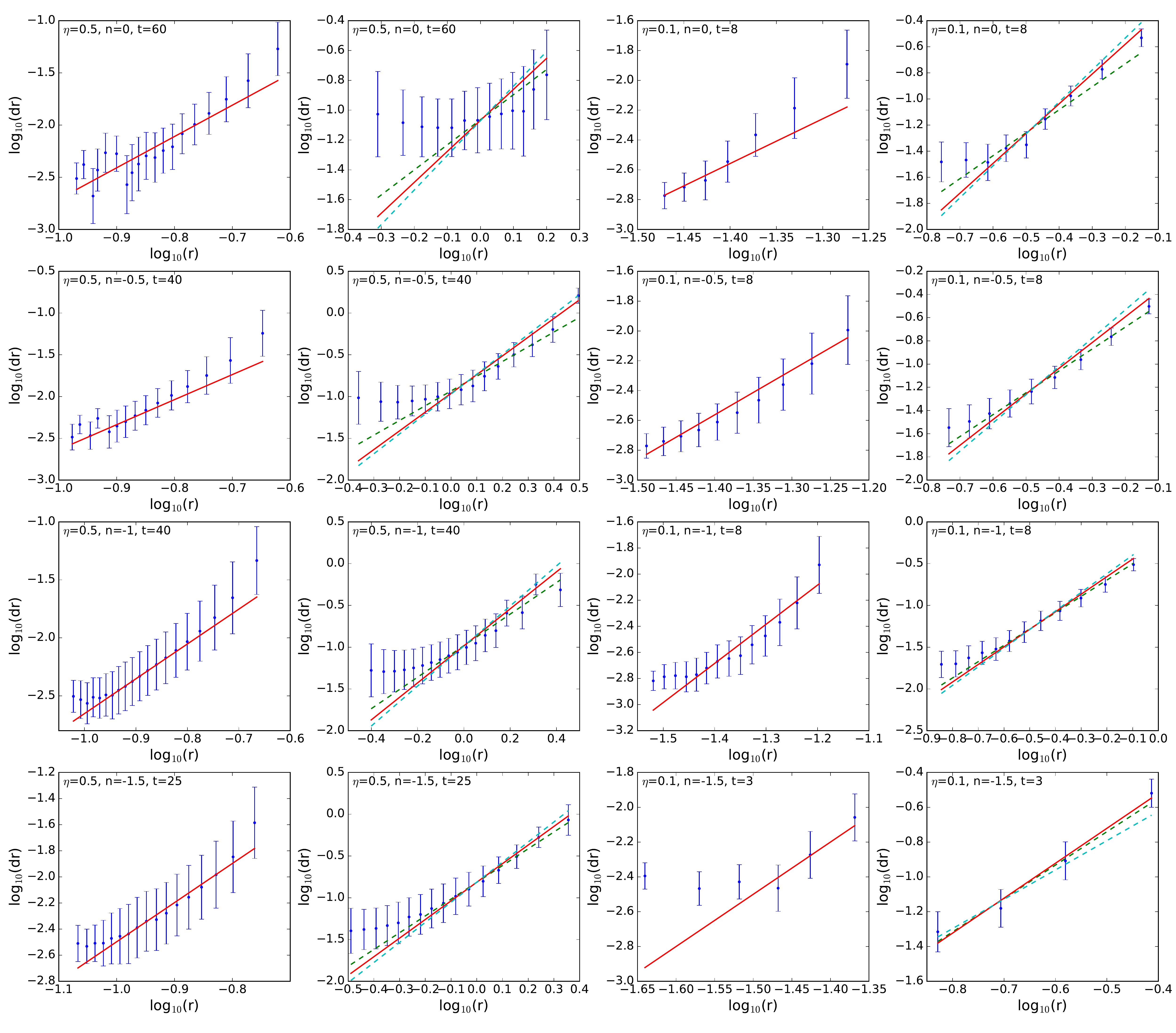} 
\caption[]{Interfold distance at null radial velocity versus self-similar predictions for a fixed value of angular momentum. The distance ${\rm d}r$ between local maxima of the function $f(r,v_r=0,j)$ is plotted as a function of $r$ for the \texttt{VlaSolve} runs with various initial conditions. The time considered corresponds to the middle column of panels in Figs.~\ref{fig:phase-space-warm} and \ref{fig:phase-space-cold}. The two lefts columns of panels correspond to the ``warm'' case, $\eta=0.5$ with $j=0.16$, and the two right ones to the ``cool'' case, $\eta=0.1$ with $j=0.06$. Then, odd column numbers (1 and 3) and even column numbers (2 and 4) correspond to the regime where centrifugal/gravitational force dominates, respectively. On each panel a red line indicates the logarithmic slope predicted by self-similarity. When the gravitational force dominates, two additional dashed curves provide a bracket of the estimated slope taking into account deviations from self-similarity, i.e. variations of the effective logarithmic slope of the gravitational force.}
\label{fig:fit_interfold}
\end{figure*}

Fig.~\ref{fig:fit_interfold} summarises the results of our measurements in the \texttt{VlaSolve} simulations. At ``small'' radius, the system is dominated by the centrifugal force, $\beta=-4$, hence ${\rm d}r \propto r^3$. This prediction is compared to measurements in the simulations in the first and third columns of Fig.~\ref{fig:fit_interfold}, which correspond respectively to the simulations with ``warm'' and ``cool'' initial conditions. At large enough radius, where the system is dominated by gravity, the force is only approximately a power-law but an average slope can nevertheless be inferred in some interval of scales $E_{\rm grav}$, corresponding to the regime where the gravitational force remains at least ten times larger than the centrifugal force and for $r$ smaller than the turnaround radius. In the second and fourth column of panels of Fig.~\ref{fig:fit_interfold}, the slope of the red line is given by the corresponding value of $1-\beta/2$. There are also two dashed cyan and green lines corresponding to the minimum and maximum value of $\beta$ found in $E_{\rm grav}$, which gives an idea of deviation from a pure power-law. Globally, the simulations agree rather well with self-similar predictions, except maybe at very small radius in the first and third columns of panels and in the top panel of the second column. Note however that measurement of the interfold distance for small values of $r$ might be partly spurious, because we are in a regime where the phase-space distribution function is small and can be affected by aliasing. Also, notice that the spiral structure survives only shortly for $(\eta,n)=(0.1,-1.5)$ which leaves only a small number of folds to deal with. Yet, the agreement with self-similarity remains good when taking into account the limitations found in all the cases at very small $r$, already after only a few dynamical times.
\begin{figure*}[htp]
\centering
\includegraphics[height=4.4cm]{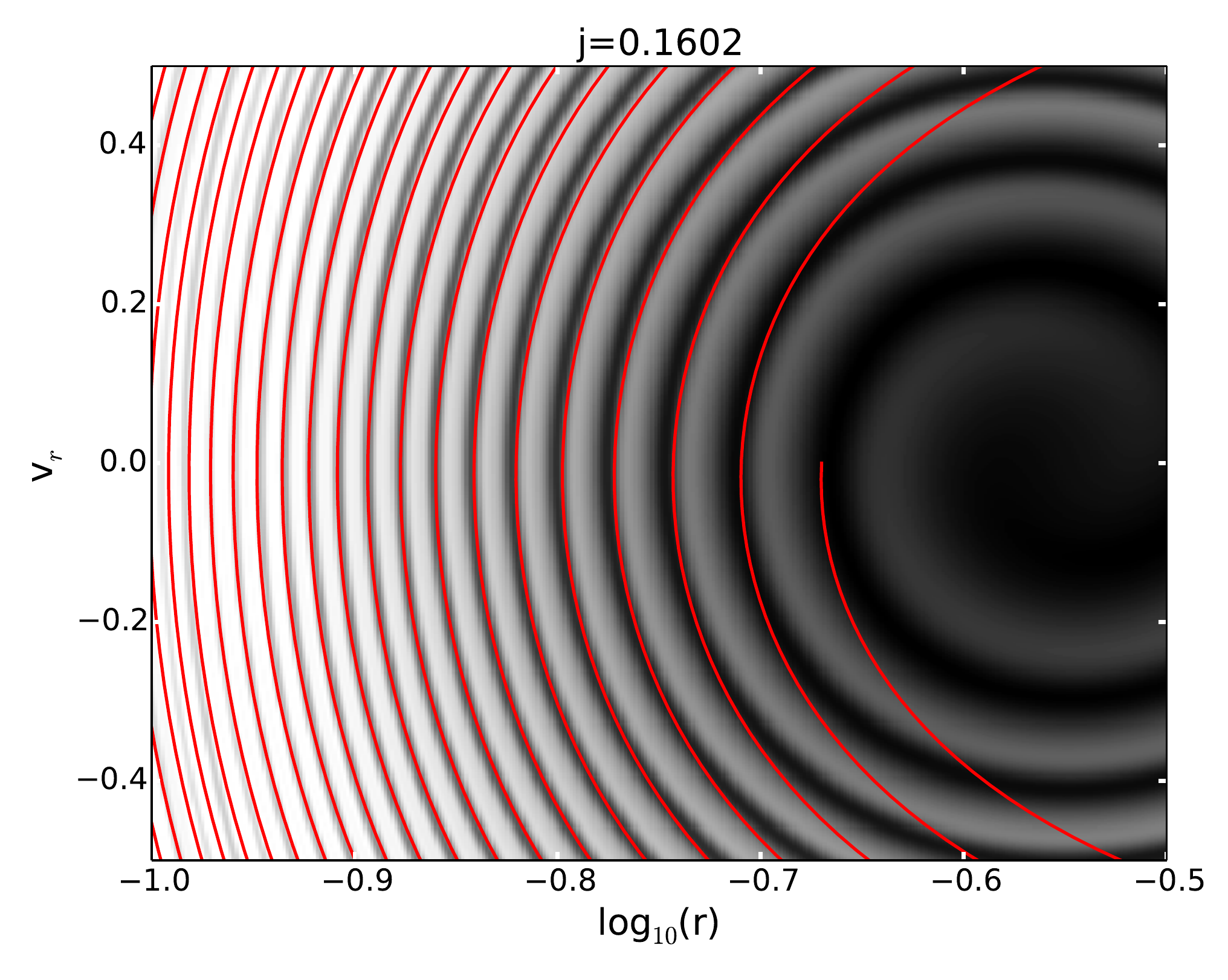}
\includegraphics[height=4.3cm]{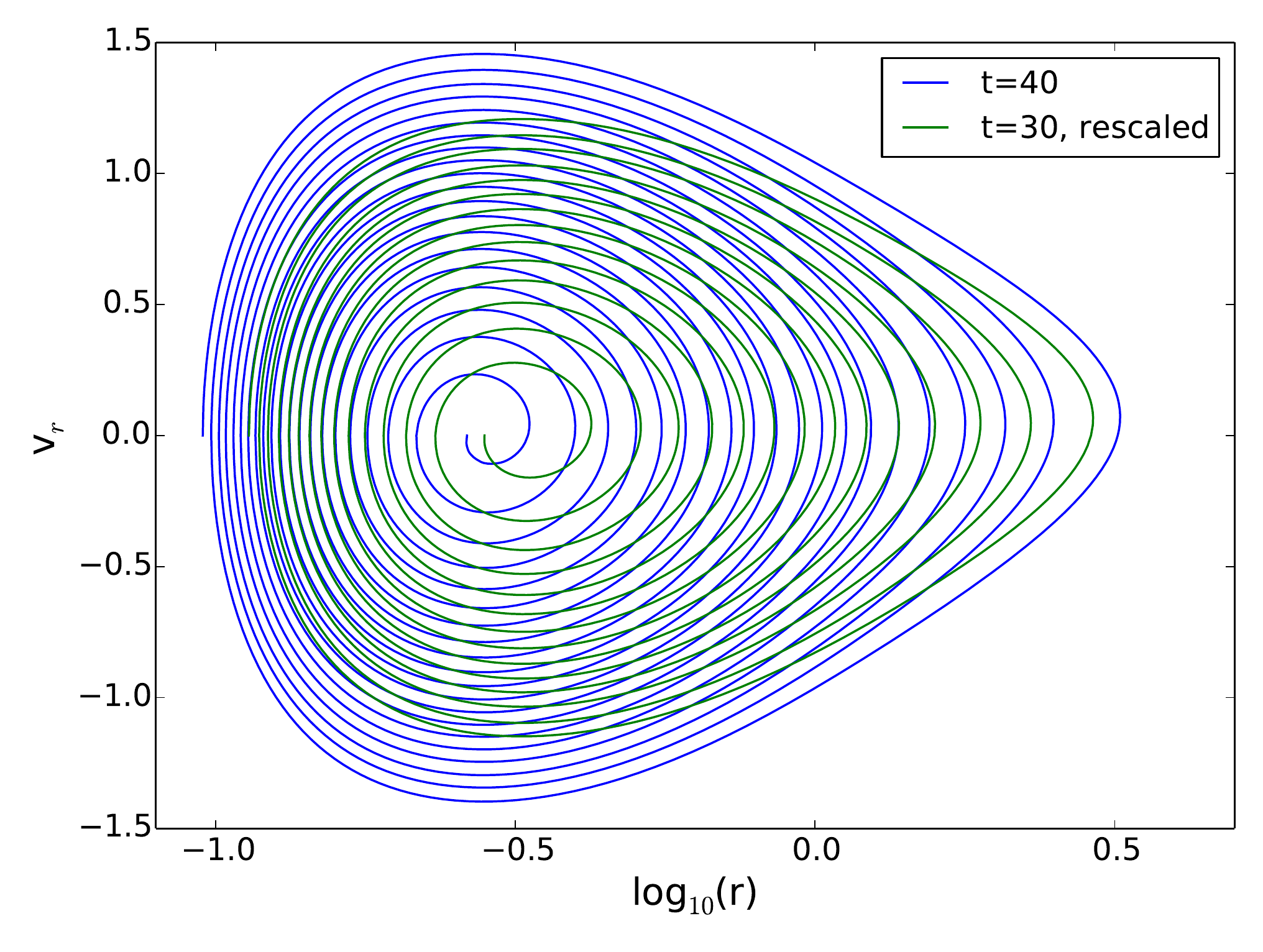}
\includegraphics[height=4.4cm]{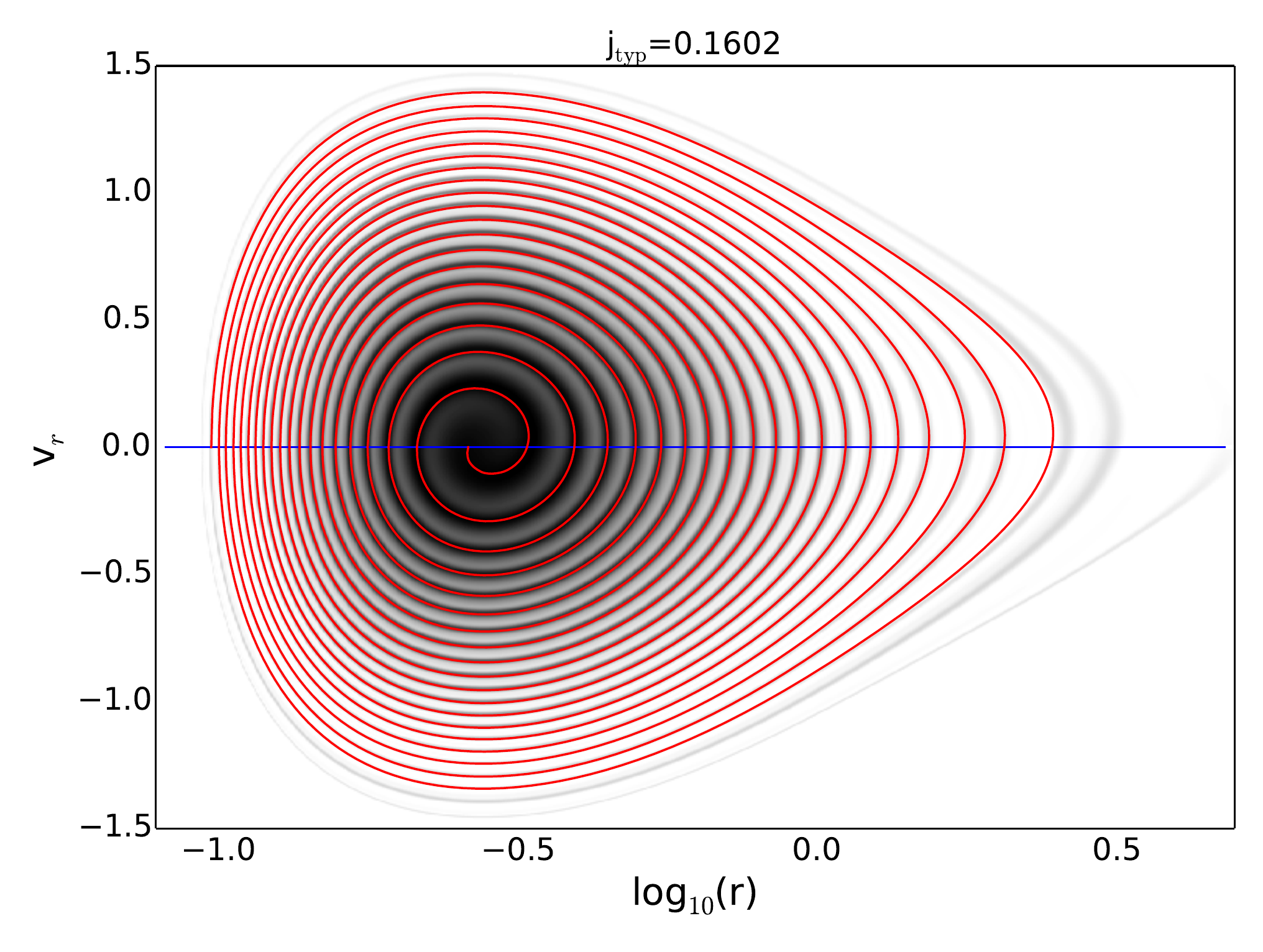}
\caption{Spiral shape versus self-similar predictions for the \texttt{VlaSolve} run with $(\eta,n)=(0.5,-1.0)$. Left panel: comparison, at $t=40$, of the local shape of the spiral structure in the region dominated by centrifugal force to the curve given by equation (\ref{eq:spi}) (red curve). Middle panel: test for self-similarity in time. The green spiral shape obtained at $t=30$, is rescaled according to equations~(\ref{eq:resca1}) and (\ref{eq:resca2}) to be compared to the blue one, in the regime dominated by centrifugal force. Right panel: using only the determination of the position of the folds in the region dominated by centrifugal force, it is possible to draw the full shape of the spiral if the gravitational potential is known, by interpolating the specific energy in Angle coordinate defined by equations~(\ref{eq:AAc1}), (\ref{eq:AAc2}) and (\ref{eq:AAc3}).}
\label{fig:spiral_selfsi}
\end{figure*}

Another interesting property than can be derived directly from equation (\ref{eq:solH}) is the local shape of the spiral near the axis $v_r=0$, hence $\Psi \simeq 0$, and for small $r$, hence small $R$. Following the unnumbered equation after equation 16 of A13 and taking into account the fact that $\alpha_2 <0$, we expect an isocontour of the function $H(R,\Psi)$ to have the following shape in the regime $R \ll 1$, $\Psi \simeq 0$,
\begin{equation}
R \propto \Psi^{-\alpha_2} \propto \Psi^{(\beta+2)/\beta}.\label{eq:spi}
\end{equation}
Note thus that because of the form of the interfold law (\ref{eq:interf}), the spiral actually locally coincides locally with a curve defined 
by $R \propto \theta^{-\alpha_2}$ with $\theta$ playing the same role as $\Psi$, but no longer restricted to $]-\pi/2,\pi/2[$, belonging instead to this interval and its multiples
$]-\pi/2+2k\pi, \pi/2 +2k\pi[$.  Left panel of figure \ref{fig:spiral_selfsi} nicely illustrates how this prediction matches well the local spiral shape of the simulated phase-space distribution function, including its local curvature, in the regime dominated by the centrifugal force ($\beta=-4$) for $(\eta,n)=(0.5,-1.0)$. 

In middle panel of this figure, we also check, in the regime dominated by centrifugal force, for self-similarity in time of the spiral shape, namely that if one considers the system at two different times, $t_1$ and $t_2$, the state at $t=t_2$ should superpose to the state at $t=t_1$ rescaled as follows
\begin{eqnarray}
v_r(t_1) & \rightarrow & v_r(t_1) \left( \frac{t_2}{t_1}\right)^{\alpha_2}, \label{eq:resca1}\\
r(t_1) & \rightarrow & r(t_1) \left( \frac{t_2}{t_1} \right)^{\alpha_1}. \label{eq:resca2}
\end{eqnarray}
Of course, since we have two distinct self-similar regimes, this property works well only in the neighbourhood of $v_r \simeq 0$ and for values of $r$ where the gravitational force is sub-dominant compared to the centrifugal force. 

Finally, right panel of Fig.~\ref{fig:spiral_selfsi} shows that if gravitational potential is known, the spiral shape of the phase-space distribution function can be fully reconstructed accurately just by knowing its intersection with the $v_r=0$ axis in the regime dominated by angular momentum (or reversely, in the regime dominated by gravitational force) by simple linear interpolation of the specific energy $E$ along the spiral during an orbit in the following Angle coordinate ${\cal A}$,
\begin{eqnarray}
{\cal A}(s,E) & \equiv & 2\pi \frac{\tau(s,E)}{T(E)}, \label{eq:AAc1} \\
\tau(s,E) &=& \int_0^s \frac{{\rm d}r(s')}{v_r(s')},  \label{eq:AAc2} \\
T(E) &=& \oint \frac{{\rm d}r(s')}{v_r(s')}, \label{eq:AAc3}
\label{eq:Angle}
\end{eqnarray}
where $s$ is a curvilinear coordinate.  This technique was actually used to draw the spiral of middle panel. Of course, this result is kind of trivial from the dynamical point of view. However, it suggests that passing to Action-Angle space or energy-Angle as performed here may represent the right way to smoothly connect both self-similar regimes and therefore to have a full analytic description of the fine grained structure of the phase-space distribution function. To do this, one needs to relate locally the angular variable $\Psi$ intervening in the self-similar solutions to the Angle given by equation (\ref{eq:Angle}), but this is left for future work. 
\section{Conclusion}
\label{sec:conclu}
In this article, we have analysed in detail the phase-space structure of various systems with spherical initial conditions, consisting in a power-law density profile with a Gaussian velocity dispersion. Two cases were considered, the ``warm'' set-up with virial ratio $\eta \simeq 0.5$ and the ``cool'' one with $\eta \simeq 0.1$. The choice of such initial conditions is not really new but the numerical set-up is different from what can be found in the literature. Firstly, we compare three kind of codes, a Vlasov code, a treecode and a shell code. Secondly we perform this comparison with unprecedented numerical resolution, namely $(N_r,N_v,N_j)=(2048,2048,128)$ for the Vlasov code and 10 million particles for the $N$-body simulations, up to 100 million for one \texttt{Gadget-2} run.

The high resolution of our simulations allowed us to study all the fine details of the phase-space distribution function and to really distinguish, for the first time, three well known dynamical phases of the evolution of these systems, namely, (a) a violent relaxation phase to a quasi-steady state where the phase-space density can be mainly described by a smooth spiral structure winding with time, (b) a steady state phase during which radial instabilities can destroy the spiral structure but do not affect the macroscopic properties of the system and (c) relaxation to a non spherical state due to radial orbit instability in the cool cases with $n \leq -1$. Obviously, we did not push the simulations long enough to approach the so-called gravothermal catastrophe regime, where a core-halo structure can appear due to collisional relaxation \citep[see, e.g.][]{Antonov1962,LyndenBell1968,LyndenBell1980,Makino1996,Baumgardt2003}.

While the concept of violent relaxation phase to a quasi-steady state is a well known process studied heavily in the literature, the fact that it is expressed as a well defined smooth spiral structure in phase-space is non trivial. Subsequent radial instabilities that can appear indeed do not introduce sufficient disorder to disturb significantly the steady state initially built from the kinematic evolution of the spiral structure. Only radial orbit instability changes the properties of the system at the coarse level. But even in this case, this happens only at small radii,  the outer part of the system being still given by the quasi-steady state solution obtained previously.

These results seem to diverge from what can be obtained in the pure cold case. For instance, a similar analysis was done by \cite{henriksen97}, but for spherical initially cold systems with power-law density profiles using a shell code: in this case, \cite{henriksen97} found that radial instabilities are sufficiently strong to destroy the quasi-steady self-similar state obtained during the violent relaxation phase and produce a density profile close to $\rho(r) \propto r^{-2}$, hence, changing the properties of the system at the macroscopic level. However, the number of shells employed by these authors was rather small and it is possible, despite the convergence tests they did, that they missed an intermediary phase where radial instability would be sufficient to destroy the spiral while preserving, as in our case, the coarse grained properties of the distribution function. Note that this question remains rather academic as radial orbit instability is expected to be prominent for such systems when allowed to deviate from spherical symmetry, although they still present some self-similar properties \citep{Vogelsberger2011}

In a second part of our analyses, in order to understand, at least partly, the dynamical processes at stake during the violent relaxation phase, we examined the properties of the spiral structure of our systems in the framework of self-similar solutions. Obviously, our systems are not fully self-similar, but we show that they follow self-similar properties in well defined domains of phase-space. Indeed, each slice of phase-space of given angular momentum can be trivially decomposed into two regions, one where centrifugal force dominates, and the other one, where gravitational force dominates. While the centrifugal force, $j^2/r^3$ is a pure power-law, this is only approximately the case for the gravitational force. Nevertheless, this approach allowed us to partly predict the properties of the spiral structure, for instance the interfold distance at zero radial velocity. While this is not enough by itself to be able to fully predict the steady state ab-initio, self-similarity in phase-space remains an interesting path of investigation. 

\begin{acknowledgements}
We thank C. Alard for providing us the main insights about the analyses performed in \S~\ref{sec:selfsi}, T. Sousbie and S. Hozumi for stimulating discussions. This work was supported in part by ANR grant ANR-13-MONU-0003. We also acknowledge the support of YITP in organising the workshop ``Vlasov-Poisson: towards numerical methods without particles'' in Kyoto, funded by grant YITP-T-15-02, ANR grant ANR-13-MONU-0003 and by Institut Lagrange de Paris (ANR-10-LABX-63 and ANR-11-IDEX-0004-02). 
\end{acknowledgements}

\bibliographystyle{aa}

\end{document}